\documentclass[5p, times, twocolumn, preprint, sort&compress]{elsarticle}

\usepackage{hyperref}
\usepackage{tabularx}
\usepackage{xcolor}
\usepackage[tbtags]{amsmath}
\allowdisplaybreaks
\usepackage{subfig}
\usepackage[none]{hyphenat}
\usepackage[perpage]{footmisc}
\usepackage{stfloats}
\usepackage{tikz}
\usepackage{pgfplots}
\usepackage{pgfplotstable}
\usepgfplotslibrary{groupplots}
\pgfplotsset{compat=1.18}
\pgfplotstableread{
hosts gpus nvms vars occ tbest tproof tprooflo tproofhi tcap gap proved runs
16 32 64 7504 0.7969 2.4400 2.4800 1.9400 3.0400 nan 0.0000 3 3
32 64 64 15904 0.3848 16.5700 16.6700 10.9200 22.4600 nan 0.0000 3 3
64 128 64 31680 0.1924 117.6000 119.7800 34.5700 119.7800 nan 0.0000 2 3
128 256 64 63232 0.0962 288.5100 826.8400 623.5800 826.8400 nan 0.0000 2 3
256 512 64 126336 0.0481 694.6200 nan nan nan 1200.0000 9.7537 0 3
}\bilpMSweep
\pgfplotstableread{
hosts gpus nvms vars occ tbest tproof tprooflo tproofhi tcap gap proved runs
32 64 8 3056 0.0449 0.3600 0.3700 0.3500 0.4000 nan 0.0000 3 3
32 64 16 4800 0.1230 0.8200 0.8600 0.6800 1.0900 nan 0.0000 3 3
32 64 32 8416 0.1895 2.3700 2.5200 2.3800 2.5900 nan 0.0000 3 3
32 64 64 15904 0.3848 16.5700 16.6700 10.9200 22.4600 nan 0.0000 3 3
32 64 128 30816 0.7559 949.8800 1011.2100 970.1400 1011.2100 nan 0.0000 2 3
}\bilpNSweep
\pgfplotstableread{
hosts gpus nvms vars occ tbest tproof tprooflo tproofhi tcap gap proved runs
2 4 13 334 0.7500 0.0300 0.0400 0.0200 0.0400 nan 0.0000 3 3
4 8 17 738 0.7656 0.0800 0.0900 0.0800 0.1000 nan 0.0000 3 3
8 16 30 2212 0.7578 0.3400 0.3600 0.3000 0.4800 nan 0.0000 3 3
16 32 57 6410 0.7539 1.8200 1.8800 1.5600 2.2000 nan 0.0000 3 3
32 64 119 29166 0.7520 295.8900 465.3000 298.4100 578.7600 nan 0.0000 3 3
64 128 242 115058 0.7510 747.0400 nan nan nan 1200.0000 2.0247 0 3
128 256 462 407836 0.7510 1133.2100 nan nan nan 1200.0000 14.2909 0 3
256 512 963 1741190 0.7502 1164.5700 nan nan nan 1200.0000 27.7050 0 3
}\bilpGrow
\newcommand{\bilpLimit}{1200}
\newcommand{\bilpLimitMin}{20}
\newcommand{\bilpMProofSixteen}{2.48}

\newcommand{\bilpNProofEight}{0.37}

\newcommand{\bilpMSteps}{%
\node[font=\tiny, anchor=south east, inner sep=1pt] at (axis cs:45.25,44.685) {\( \times 7.2 \)};
\node[font=\tiny, anchor=south east, inner sep=1pt] at (axis cs:90.51,314.704) {\( \times 6.9 \)};
}

\newcommand{\bilpNSteps}{%
\node[font=\tiny, anchor=south east, inner sep=1pt] at (axis cs:45.25,6.481) {\( \times 6.6 \)};
\node[font=\tiny, anchor=south east, inner sep=1pt] at (axis cs:90.51,129.834) {\( \times 61 \)};
}
\newcommand{\bilpNStepA}{6.6}
\newcommand{\bilpNStepB}{61}
\newcommand{\bilpNProofOneTwentyEightMin}{17}
\newcommand{\bilpBeyondProved}{2}
\newcommand{\bilpBeyondRuns}{15}

\pgfplotstableread{
load energy energy_ci epw epw_ci switch switch_ci acc acc_ci reqacc reqacc_ci pm pm_ci pmabs pmabs_ci gpu gpu_ci pack pack_ci cf cf_ci wait wait_ci waitp waitp_ci mig mig_ci inter inter_ci gbh gbh_ci copy copy_ci migvm migvm_ci calls calls_ci callmed callmed_ci callp95 callp95_ci wall wall_ci n
0.5 3.7407 0.3529 0.4422 0.0067 3.8361 0.1365 100.0000 0.0000 100.0000 0.0000 58.1261 4.7274 74.4014 6.0511 56.1509 4.6485 82.9888 0.5593 13.3022 0.5904 0.0000 0.0000 0.0000 0.0000 0.0000 0.0000 0.0000 0.0000 0.0000 0.0000 0.0000 0.0000 0.0000 0.0000 0.0000 0.0000 0.0000 0.0000 0.0000 0.0000 0.0016 0.0002 5
0.65 4.8719 0.3245 0.4407 0.0047 4.7806 0.1315 99.8898 0.3059 99.9486 0.1427 75.3911 4.0632 96.5006 5.2009 73.0168 4.0560 83.5006 0.6689 12.9016 0.8146 0.0000 0.0000 0.0000 0.0000 0.0000 0.0000 0.0000 0.0000 0.0000 0.0000 0.0000 0.0000 0.0000 0.0000 0.0000 0.0000 0.0000 0.0000 0.0000 0.0000 0.0021 0.0002 5
0.75 5.5691 0.2317 0.4399 0.0021 5.1889 0.3488 98.4604 0.9136 99.2486 0.4470 85.5598 3.0620 109.5166 3.9194 83.1122 2.8638 84.0156 0.4564 12.5298 0.3282 0.0000 0.0000 0.0000 0.0000 0.0000 0.0000 0.0000 0.0000 0.0000 0.0000 0.0000 0.0000 0.0000 0.0000 0.0000 0.0000 0.0000 0.0000 0.0000 0.0000 0.0025 0.0003 5
0.85 6.0811 0.0990 0.4389 0.0021 5.2639 0.5505 94.5686 1.4041 97.2456 0.7928 92.8058 0.6707 118.7914 0.8585 90.3052 0.5375 84.6490 0.8218 11.8720 1.0327 0.0000 0.0000 0.0000 0.0000 0.0000 0.0000 0.0000 0.0000 0.0000 0.0000 0.0000 0.0000 0.0000 0.0000 0.0000 0.0000 0.0000 0.0000 0.0000 0.0000 0.0031 0.0005 5
1 6.5189 0.0983 0.4361 0.0020 4.0917 0.5468 83.7090 3.2223 91.7240 1.6987 98.0017 0.5721 125.4422 0.7323 95.7651 0.7230 86.1708 0.7734 10.3416 0.9291 0.0000 0.0000 0.0000 0.0000 0.0000 0.0000 0.0000 0.0000 0.0000 0.0000 0.0000 0.0000 0.0000 0.0000 0.0000 0.0000 0.0000 0.0000 0.0000 0.0000 0.0039 0.0006 5
1.2 6.7369 0.0387 0.4318 0.0009 2.3639 0.7972 72.7824 3.2753 85.7362 1.9222 99.2972 0.3601 127.1004 0.4610 97.5412 0.4291 88.6062 0.5285 7.8788 0.6776 0.0000 0.0000 0.0000 0.0000 0.0000 0.0000 0.0000 0.0000 0.0000 0.0000 0.0000 0.0000 0.0000 0.0000 0.0000 0.0000 0.0000 0.0000 0.0000 0.0000 0.0050 0.0007 5
}{\evalSoneFF}
\pgfplotstableread{
load energy energy_ci epw epw_ci switch switch_ci acc acc_ci reqacc reqacc_ci pm pm_ci pmabs pmabs_ci gpu gpu_ci pack pack_ci cf cf_ci wait wait_ci waitp waitp_ci mig mig_ci inter inter_ci gbh gbh_ci copy copy_ci migvm migvm_ci calls calls_ci callmed callmed_ci callp95 callp95_ci wall wall_ci n
0.5 3.7201 0.3377 0.4398 0.0054 2.2139 0.2720 100.0000 0.0000 100.0000 0.0000 56.5197 3.9083 72.3452 5.0027 52.0339 3.6942 89.5150 0.5776 7.9076 0.5411 0.0000 0.0000 0.0000 0.0000 0.0000 0.0000 0.0000 0.0000 0.0000 0.0000 0.0000 0.0000 0.0000 0.0000 0.0000 0.0000 0.0000 0.0000 0.0000 0.0000 0.0013 0.0002 5
0.65 4.8482 0.3185 0.4384 0.0037 2.7000 0.2963 99.9632 0.1022 99.9794 0.0572 73.4316 3.7332 93.9924 4.7785 68.1924 3.4027 89.4324 1.0410 8.0898 0.9680 0.0000 0.0000 0.0000 0.0000 0.0000 0.0000 0.0000 0.0000 0.0000 0.0000 0.0000 0.0000 0.0000 0.0000 0.0000 0.0000 0.0000 0.0000 0.0000 0.0000 0.0015 0.0002 5
0.75 5.5704 0.2508 0.4363 0.0037 2.8139 0.4074 99.3844 0.5423 99.7092 0.2531 83.5516 3.2399 106.9460 4.1470 78.1560 2.9784 90.0850 0.7269 7.4916 0.6423 0.0000 0.0000 0.0000 0.0000 0.0000 0.0000 0.0000 0.0000 0.0000 0.0000 0.0000 0.0000 0.0000 0.0000 0.0000 0.0000 0.0000 0.0000 0.0000 0.0000 0.0017 0.0004 5
0.85 6.1624 0.1459 0.4358 0.0028 2.8500 0.3936 96.6766 0.3302 98.4164 0.2078 91.9411 1.6375 117.6846 2.0960 86.0416 1.4848 90.6172 1.1954 6.9700 1.0571 0.0000 0.0000 0.0000 0.0000 0.0000 0.0000 0.0000 0.0000 0.0000 0.0000 0.0000 0.0000 0.0000 0.0000 0.0000 0.0000 0.0000 0.0000 0.0000 0.0000 0.0019 0.0004 5
1 6.6749 0.1063 0.4324 0.0022 2.0639 0.2057 86.9686 2.7019 93.6872 1.4688 98.0864 0.6423 125.5506 0.8221 93.0732 1.3312 91.5494 0.3301 6.0412 0.4069 0.0000 0.0000 0.0000 0.0000 0.0000 0.0000 0.0000 0.0000 0.0000 0.0000 0.0000 0.0000 0.0000 0.0000 0.0000 0.0000 0.0000 0.0000 0.0000 0.0000 0.0021 0.0005 5
1.2 6.9080 0.0489 0.4288 0.0017 1.1639 0.2362 74.9466 2.5422 87.3506 1.5259 99.4655 0.2402 127.3158 0.3075 96.0406 0.7036 92.7818 0.4773 4.7376 0.4079 0.0000 0.0000 0.0000 0.0000 0.0000 0.0000 0.0000 0.0000 0.0000 0.0000 0.0000 0.0000 0.0000 0.0000 0.0000 0.0000 0.0000 0.0000 0.0000 0.0000 0.0025 0.0005 5
}{\evalSoneMFI}
\pgfplotstableread{
load energy energy_ci epw epw_ci switch switch_ci acc acc_ci reqacc reqacc_ci pm pm_ci pmabs pmabs_ci gpu gpu_ci pack pack_ci cf cf_ci wait wait_ci waitp waitp_ci mig mig_ci inter inter_ci gbh gbh_ci copy copy_ci migvm migvm_ci calls calls_ci callmed callmed_ci callp95 callp95_ci wall wall_ci n
0.5 3.6429 0.3511 0.4306 0.0082 3.6000 0.1525 100.0000 0.0000 100.0000 0.0000 52.8220 4.4009 67.6122 5.6332 48.1367 3.6415 96.7770 0.2396 1.7414 0.1613 0.0000 0.0000 0.0000 0.0000 12.7194 0.8358 12.3722 0.8155 725.3944 35.0245 37.5640 1.7357 0.5976 0.0408 8.0000 0.0000 0.0047 0.0006 0.0142 0.0039 0.0028 0.0003 5
0.65 4.7344 0.3209 0.4280 0.0044 4.2833 0.3476 100.0000 0.0000 100.0000 0.0000 67.9998 3.8046 87.0398 4.8699 62.8067 3.6844 97.1320 0.1352 1.5198 0.0984 0.0000 0.0000 0.0000 0.0000 15.4444 1.0673 15.0444 1.0582 875.8806 62.5983 37.2853 2.0090 0.5660 0.0385 8.0000 0.0000 0.0063 0.0011 0.0192 0.0029 0.0037 0.0006 5
0.75 5.4745 0.2529 0.4273 0.0040 4.7917 0.2066 99.7336 0.4459 99.8682 0.2134 78.3980 3.1863 100.3494 4.0784 72.6241 2.9438 97.2684 0.1747 1.4344 0.0692 0.0000 0.0000 0.0000 0.0000 17.2694 0.9372 16.9556 1.0045 976.2444 51.4527 36.8570 0.3235 0.5495 0.0230 8.0000 0.0000 0.0069 0.0008 0.0192 0.0037 0.0040 0.0006 5
0.85 6.1333 0.1668 0.4285 0.0028 5.1417 0.4485 98.2350 0.7838 99.1454 0.3733 87.9959 2.0047 112.6348 2.5660 81.0056 1.9481 97.4204 0.2240 1.2918 0.0660 0.0000 0.0000 0.0000 0.0000 19.5417 1.3536 19.2333 1.4222 1069.4111 52.4241 35.6269 1.2547 0.5560 0.0349 8.0000 0.0000 0.0078 0.0006 0.0211 0.0031 0.0046 0.0005 5
1 6.7647 0.1105 0.4276 0.0015 4.3361 0.5869 90.7820 2.9230 95.4420 1.6229 96.3300 1.1309 123.3024 1.4476 89.5546 1.3867 97.4762 0.0926 1.2304 0.0378 0.0000 0.0000 0.0000 0.0000 20.1639 1.5210 19.9167 1.5148 1088.6278 74.9725 34.9933 0.4629 0.5027 0.0323 8.0000 0.0000 0.0085 0.0008 0.0238 0.0020 0.0052 0.0008 5
1.2 7.0427 0.0622 0.4256 0.0013 2.6972 0.6655 77.2892 3.5830 88.3514 2.0880 99.1583 0.3287 126.9226 0.4207 93.7472 0.9059 97.5702 0.1829 1.1304 0.0817 0.0000 0.0000 0.0000 0.0000 20.3472 1.3969 20.0806 1.3506 1039.8194 73.4980 33.1453 1.0374 0.4544 0.0266 8.0000 0.0000 0.0091 0.0006 0.0242 0.0034 0.0053 0.0005 5
}{\evalSoneTC}
\pgfplotstableread{
load energy energy_ci epw epw_ci switch switch_ci acc acc_ci reqacc reqacc_ci pm pm_ci pmabs pmabs_ci gpu gpu_ci pack pack_ci cf cf_ci wait wait_ci waitp waitp_ci mig mig_ci inter inter_ci gbh gbh_ci copy copy_ci migvm migvm_ci calls calls_ci callmed callmed_ci callp95 callp95_ci wall wall_ci n
0.5 3.5894 0.3406 0.4243 0.0068 7.5222 0.3209 100.0000 0.0000 100.0000 0.0000 50.8420 4.0787 65.0778 5.2207 50.3837 4.0168 92.4762 0.4057 5.0750 0.3231 0.0000 0.0000 0.0000 0.0000 9.0694 0.9857 8.5583 0.8930 919.1389 83.3395 68.7981 2.2667 0.4263 0.0514 28.3306 1.0479 3.2928 0.8551 8.1666 2.1377 2.7928 0.4703 5
0.65 4.6881 0.3283 0.4238 0.0044 9.1694 0.4619 99.9632 0.1022 99.9794 0.0572 66.5003 4.2656 85.1204 5.4599 65.8973 4.1212 92.5778 0.6150 5.0896 0.5432 0.0100 0.0176 0.0000 0.0000 10.6222 1.1583 10.0028 1.0862 1110.6111 97.6298 71.1546 2.0865 0.3893 0.0415 35.6639 0.7858 4.4932 0.7170 10.0290 0.1218 4.4512 0.6659 5
0.75 5.4349 0.2650 0.4244 0.0037 9.8861 0.6906 99.5444 0.7316 99.7616 0.3811 77.3739 3.8572 99.0386 4.9372 76.6831 3.7478 92.0936 0.8403 5.4634 0.6703 0.0720 0.0442 0.0000 0.0000 11.3917 1.0808 10.7556 1.1201 1210.3250 143.1410 72.0061 3.6740 0.3629 0.0328 40.1528 0.8276 5.0446 0.7986 10.1309 0.0471 5.5985 0.8492 5
0.85 6.1005 0.2109 0.4240 0.0028 10.5278 0.9363 98.9234 1.0278 99.3944 0.5590 86.4786 3.0501 110.6926 3.9041 85.7534 2.8911 92.5042 0.6439 5.0910 0.3839 0.3740 0.1298 0.0000 0.0000 12.0222 1.8522 11.3806 1.7029 1275.5528 163.6182 71.9025 3.8075 0.3410 0.0490 44.3583 1.1413 5.2128 1.5439 10.1533 0.0348 6.5774 0.8580 5
1 6.7484 0.1384 0.4235 0.0017 8.7444 1.2151 91.4820 2.7453 95.1674 1.7161 95.2991 1.6551 121.9828 2.1185 94.5708 1.5682 92.9402 0.5765 4.6284 0.4548 1.5140 0.7132 0.0000 0.0000 11.5361 1.2698 10.7917 1.2513 1137.0972 180.1812 67.2963 4.4710 0.2888 0.0357 49.1528 1.0592 4.8527 2.3207 10.1784 0.0653 7.1630 1.2175 5
1.2 7.0401 0.0356 0.4220 0.0005 5.4194 0.9037 77.4072 3.8217 86.5488 2.7850 98.7237 0.2885 126.3664 0.3693 98.1301 0.3056 93.9066 0.4209 3.8090 0.3660 2.9200 0.4528 31.3160 3.2528 10.4139 0.4873 9.3833 0.4701 888.8167 89.4758 60.5731 3.9472 0.2375 0.0092 55.8917 0.9941 0.4792 0.8127 10.1399 0.0856 6.2346 1.3399 5
}{\evalSoneDMIG}

\newcommand{\evalSoneGainMFIDpmZeropFive}{2.7}

\newcommand{\evalSoneTCPackZeropFive}{96.8}

\newcommand{\evalSoneTCMigZeropFive}{12.7}

\newcommand{\evalSoneGainTCDpmZeropFive}{9.1}

\newcommand{\evalSoneDMIGPackZeropFive}{92.5}

\newcommand{\evalSoneDMIGMigZeropFive}{9.1}

\newcommand{\evalSoneGainDMIGDpmZeropFive}{12.5}

\newcommand{\evalSoneGainDMIGDepwZeropFive}{4.0}

\newcommand{\evalSoneGainTCDpmZeropSixFive}{9.8}

\newcommand{\evalSoneGainDMIGDpmZeropSixFive}{11.8}

\newcommand{\evalSoneGainDMIGDpmZeropEightFive}{6.8}

\newcommand{\evalSoneFFAccOne}{83.7}

\newcommand{\evalSoneMFIAccOne}{87.0}

\newcommand{\evalSoneTCAccOne}{90.8}

\newcommand{\evalSoneTCMigOne}{20.2}

\newcommand{\evalSoneDMIGAccOne}{91.5}

\newcommand{\evalSoneDMIGMigOne}{11.5}
\newcommand{\evalSoneDMIGWaitOne}{1.5}
\newcommand{\evalSoneGainDMIGDaccOne}{7.8}
\newcommand{\evalSoneGainDMIGDpmOne}{2.8}

\newcommand{\evalSoneGainDMIGDepwOne}{2.9}

\newcommand{\evalSoneVsTCDpmMax}{3.7}
\newcommand{\evalSoneVsTCDpmMin}{0.4}
\newcommand{\evalSoneVsTCDmigMin}{29}
\newcommand{\evalSoneVsTCDmigMax}{49}
\newcommand{\evalSoneVsTCWins}{29}
\newcommand{\evalSoneVsTCRuns}{30}
\newcommand{\evalSoneSeeds}{five}
\newcommand{\evalSoneRows}{%
			0.75 & FF & 98.5 & 109.5 & 83.1 & 84.0 & 5.57 & 5.2 & 0 & 0 \\
			 & MFI & 99.4 & 106.9 & 78.2 & 90.1 & 5.57 & 2.8 & 0 & 0 \\
			 & TC & 99.7 & 100.3 & 72.6 & 97.3 & 5.47 & 4.8 & 17.3 & 0 \\
			 & OrigaMIG & 99.5 & 99.0 & 76.7 & 92.1 & 5.43 & 9.9 & 11.4 & 0.1 \\ \hline
			1.0 & FF & 83.7 & 125.4 & 95.8 & 86.2 & 6.52 & 4.1 & 0 & 0 \\
			 & MFI & 87.0 & 125.6 & 93.1 & 91.5 & 6.67 & 2.1 & 0 & 0 \\
			 & TC & 90.8 & 123.3 & 89.6 & 97.5 & 6.76 & 4.3 & 20.2 & 0 \\
			 & OrigaMIG & 91.5 & 122.0 & 94.6 & 92.9 & 6.75 & 8.7 & 11.5 & 1.5 \\ \hline}

\pgfplotstableread{
v dacc dacc_ci dpm dpm_ci dgpu dgpu_ci dpack dpack_ci denergy denergy_ci depw depw_ci n
0 1.9830 2.5559 0.5443 2.5741 4.1959 1.9919 5.7537 1.0324 -1.6780 1.7458 0.6753 1.3213 3
1 -0.1947 1.2504 -0.6100 0.8675 1.4996 2.8819 1.6167 1.8007 -0.3173 0.6582 0.0004 0.3792 3
2 1.8440 4.8004 2.0845 3.9346 3.2802 4.3888 4.2500 2.7878 -0.5185 3.9177 0.9840 0.9653 3
3 0.7047 0.7801 3.2688 1.5130 3.5225 0.8756 3.7507 0.7025 0.5616 0.8459 1.0486 0.7605 3
4 3.5447 4.9948 3.0351 2.0478 4.5168 2.0587 5.7357 2.6091 -0.7668 0.9324 1.4508 1.2826 3
5 0.3570 0.8631 2.9161 4.3661 6.0263 5.5774 5.6037 5.5626 0.5115 1.3901 0.7562 1.6691 3
6 2.6280 0.9596 0.0390 1.2094 3.6046 1.8343 5.5680 1.4225 -2.1288 0.9278 0.6457 0.1539 3
7 2.6150 1.3900 1.1912 0.7928 4.8131 3.4485 6.0933 0.8738 -1.4293 2.5850 0.6001 0.5514 3
8 2.4533 0.8097 0.9588 3.5972 5.0320 3.7898 6.2823 1.8493 -1.5709 2.4226 0.4425 0.5465 3
9 1.9977 3.5701 4.0446 5.5599 7.0546 7.0190 7.8620 4.3332 1.0219 3.8986 2.3523 1.5549 3
10 -0.6340 1.1975 -0.4243 0.7360 8.9416 2.7173 6.8097 1.9682 0.5009 1.7191 0.4652 0.5180 3
11 0.2457 0.3104 0.4363 0.2966 0.9175 0.4066 0.9277 0.4295 0.0257 0.0540 0.0619 0.0714 3
}{\evalStwoGainMFI}

\newcommand{\evalStwoTCDpmMixSmall}{2.7}
\newcommand{\evalStwoTCMigMixSmall}{29.9}

\newcommand{\evalStwoTCDpmMixLarge}{4.5}

\newcommand{\evalStwoTCDaccTwoGi}{5.3}
\newcommand{\evalStwoTCDpmTwoGi}{7.8}

\newcommand{\evalStwoTCDpmLognormal}{9.4}

\newcommand{\evalStwoTCDaccTenantLow}{5.4}

\newcommand{\evalStwoTCDpmAhundred}{5.7}
\newcommand{\evalStwoTCMigAhundred}{18.5}

\newcommand{\evalStwoTCDpmAthirty}{6.0}

\newcommand{\evalStwoTCDpmCpu}{0.7}
\newcommand{\evalStwoTCMigCpu}{18.1}

\newcommand{\evalStwoTCDpmTrace}{1.4}
\newcommand{\evalStwoTCMigTrace}{10.7}
\newcommand{\evalStwoTCGbhTrace}{615}
\pgfplotstableread{
v dacc dacc_ci dpm dpm_ci dgpu dgpu_ci dpack dpack_ci denergy denergy_ci depw depw_ci n
0 3.9137 3.5825 4.8225 3.0872 9.9384 2.2449 12.6267 0.9795 -1.2148 2.0572 2.2830 1.4706 3
1 0.4030 0.9369 2.6542 0.8078 8.0985 2.6255 8.4017 1.7330 0.2905 0.3327 0.9122 0.8772 3
2 3.2640 6.1598 4.5248 6.9826 6.4644 6.2100 8.1450 3.8233 -0.4471 6.8373 1.9340 1.4902 3
3 0.8937 1.4083 5.5011 3.2695 7.6597 1.1277 7.9573 0.8288 1.1054 1.3613 1.5718 1.1984 3
4 5.2527 7.3639 7.7645 3.6861 9.3817 4.4456 12.1233 3.2637 -0.4649 3.4691 3.5893 1.1185 3
5 0.4407 1.0870 9.4478 1.9482 13.6999 0.9955 13.4963 0.3042 2.5855 0.9563 2.8473 0.1867 3
6 5.3870 1.0459 3.3167 1.1294 7.9948 1.5664 12.1437 0.5663 -2.8075 2.0476 2.3817 0.9624 3
7 4.2357 4.8185 6.4420 9.0041 10.3024 6.2928 12.8763 1.5615 -0.7206 7.4732 2.5665 2.1255 3
8 3.6037 1.6772 5.6928 3.4616 10.6705 3.6831 12.6673 1.9941 -0.7628 3.1089 1.9113 0.3909 3
9 2.7183 4.9848 6.0005 4.2934 11.0734 6.9212 12.4603 3.7315 1.6866 3.7893 3.3270 1.1389 3
10 -1.5557 2.6503 0.7030 1.5451 21.9692 3.5714 18.9497 2.4603 2.3100 2.1274 1.2698 0.4835 3
11 0.3697 0.1849 1.4498 0.7725 2.0700 0.2657 2.0950 0.2863 0.2605 0.1895 0.3187 0.2233 3
}{\evalStwoGainTC}

\newcommand{\evalStwoDMIGDpmMixSmall}{4.1}
\newcommand{\evalStwoDMIGMigMixSmall}{9.8}

\newcommand{\evalStwoDMIGDpmMixLarge}{2.3}

\newcommand{\evalStwoDMIGDpmTwoGi}{4.7}

\newcommand{\evalStwoDMIGDpmLognormal}{11.2}

\newcommand{\evalStwoDMIGDpmAhundred}{5.6}
\newcommand{\evalStwoDMIGMigAhundred}{9.3}

\newcommand{\evalStwoDMIGDpmAthirty}{3.4}

\newcommand{\evalStwoDMIGDaccCpu}{2.6}
\newcommand{\evalStwoDMIGDpmCpu}{2.6}
\newcommand{\evalStwoDMIGMigCpu}{9.4}

\newcommand{\evalStwoDMIGDpmTrace}{7.5}
\newcommand{\evalStwoDMIGMigTrace}{61.3}
\newcommand{\evalStwoDMIGGbhTrace}{8772}
\pgfplotstableread{
v dacc dacc_ci dpm dpm_ci dgpu dgpu_ci dpack dpack_ci denergy denergy_ci depw depw_ci n
0 4.8253 1.7157 6.4691 4.5642 4.6830 4.8156 7.8353 2.7358 -0.7551 2.9605 3.3032 1.1785 3
1 0.8953 0.3641 4.1045 2.0482 2.7977 1.8351 3.5810 1.1998 0.7914 0.6715 1.9567 0.1688 3
2 2.3547 3.7626 2.3220 1.7275 0.9083 2.7343 2.3230 4.3970 -0.3779 3.3184 1.3461 1.4547 3
3 0.7263 1.3179 6.1742 1.2388 3.1075 1.2020 3.2773 1.5299 1.6488 0.5473 2.0535 0.6867 3
4 4.5293 5.0004 4.7015 3.2460 3.9391 4.0535 5.9343 4.1155 -0.5319 1.3597 2.5210 1.1682 3
5 0.4640 1.1744 11.1942 0.8215 9.5874 0.9933 9.0893 0.7076 3.4568 0.5752 3.7163 0.3235 3
6 6.0967 0.9358 5.1186 1.1310 3.3944 1.2569 7.9603 0.4831 -2.1920 2.1344 3.4269 0.7736 3
7 4.0047 4.2276 7.3093 6.5119 5.2347 6.0459 7.6033 1.1410 -0.0458 6.4870 3.1386 1.4038 3
8 3.2317 2.5756 5.5960 6.6278 4.7365 6.4970 6.3403 5.9528 -0.1424 3.1248 2.2080 1.8544 3
9 2.3043 4.1912 3.4383 5.4954 2.9846 4.7385 3.8340 4.3320 0.7671 3.1491 2.0296 2.2811 3
10 2.6073 1.7546 2.5888 1.7208 -3.6598 4.4932 0.8867 1.0933 -5.5608 2.8693 -0.2716 0.7673 3
11 0.3497 0.2649 7.5452 1.1308 0.7593 0.1375 0.7607 0.1787 2.0630 0.2905 2.0881 0.3497 3
}{\evalStwoGainDMIG}

\newcommand{\evalStwoFFAccTwoGi}{92.4}

\newcommand{\evalStwoFFAccTenantLow}{91.9}

\newcommand{\evalStwoFFAccCpu}{70.3}

\newcommand{\evalSthreeTCQuarterMig}{19.9}

\newcommand{\evalSthreeTCQuarterDpm}{4.8}

\newcommand{\evalSthreeTCHourMig}{10.0}

\newcommand{\evalSthreeTCHourDpm}{3.1}

\newcommand{\evalSthreeDeltaZeroMig}{142.5}
\newcommand{\evalSthreeDeltaZeroMigvm}{4.05}
\newcommand{\evalSthreeDeltaZeroGbh}{21106}

\newcommand{\evalSthreeDeltaZeroDpm}{8.0}

\newcommand{\evalSthreeDeltaFiveMig}{12.3}

\newcommand{\evalSthreeDeltaFiveDpm}{6.3}

\newcommand{\evalSthreeDeltaTenMig}{12.8}
\newcommand{\evalSthreeDeltaTenMigvm}{0.36}
\newcommand{\evalSthreeDeltaTenGbh}{1316}
\newcommand{\evalSthreeDeltaTenCopy}{70}
\newcommand{\evalSthreeDeltaTenDpm}{6.5}

\newcommand{\evalSthreeDeltaTwentyMig}{6.2}

\newcommand{\evalSthreeDeltaTwentyDpm}{4.3}

\newcommand{\evalSthreeRows}{%
			TC, every 15\,min & 3 & 19.9 & 0.56 & 1085 & 98.55 & 113.4 & 6.19 & -- \\ \hline
			TC, every hour & 3 & 10.0 & 0.28 & 530 & 98.04 & 115.4 & 6.22 & -- \\ \hline
			OrigaMIG, \( \delta_i = 0 \) & 3 & 142.5 & 4.05 & 21106 & 98.68 & 109.6 & 6.14 & 4.10 \\ \hline
			OrigaMIG, \( \delta_i = 5 \) & 3 & 12.3 & 0.35 & 1272 & 99.46 & 111.6 & 6.17 & 0.35 \\ \hline
			OrigaMIG, \( \delta_i = 10 \) & 3 & 12.8 & 0.36 & 1316 & 99.46 & 111.4 & 6.16 & 5.64 \\ \hline
			OrigaMIG, \( \delta_i = 20 \) & 3 & 6.2 & 0.18 & 713 & 97.33 & 114.0 & 6.16 & 4.98 \\ \hline
			OrigaMIG, \( \delta_i = 40 \) & 3 & 0.1 & 0.00 & 2 & 94.66 & 119.2 & 6.13 & 4.92 \\ \hline}
\pgfplotstableread{
delta energy energy_ci epw epw_ci switch switch_ci acc acc_ci reqacc reqacc_ci pm pm_ci pmabs pmabs_ci gpu gpu_ci pack pack_ci cf cf_ci wait wait_ci waitp waitp_ci mig mig_ci inter inter_ci gbh gbh_ci copy copy_ci migvm migvm_ci calls calls_ci callmed callmed_ci callp95 callp95_ci wall wall_ci dacc dacc_ci dpm dpm_ci dgpu dgpu_ci dpack dpack_ci denergy denergy_ci depw depw_ci n
0 6.1362 0.3490 0.4260 0.0047 51.8889 32.0760 98.6840 1.3794 99.0997 0.7733 85.6484 5.7201 109.6300 7.3217 84.0436 5.2311 94.4380 2.2261 4.0077 1.3956 21.0233 67.2765 0.0000 0.0000 142.5231 109.0040 135.8565 105.1712 21105.8657 15120.6403 99.8911 7.3530 4.0529 3.1747 30.2361 5.3947 4.0988 12.9111 10.0932 0.0530 5.0282 3.7232 4.0523 1.8951 7.9890 5.1356 7.1968 5.4994 9.7603 3.2709 -0.3105 3.7776 3.1300 1.3588 3
5 6.1720 0.2643 0.4256 0.0046 10.6343 0.8348 99.4563 1.3220 99.6993 0.7468 87.1953 2.8854 111.6100 3.6933 86.4949 3.0022 92.3540 1.1535 5.0490 0.6774 0.6233 0.0574 0.0000 0.0000 12.3102 1.8096 11.8519 1.6071 1272.3241 201.0124 68.6831 1.9688 0.3475 0.0543 44.4630 0.7108 0.3470 0.2287 5.4411 2.7460 1.7193 0.4863 4.8247 1.5455 6.3240 1.9135 4.4898 2.8268 7.6763 0.8965 -0.8997 2.6930 3.2161 0.7637 3
10 6.1632 0.2697 0.4252 0.0039 10.8472 1.4813 99.4570 1.2744 99.6863 0.6784 87.0635 5.3438 111.4413 6.8401 86.3211 4.8124 92.5130 1.4846 5.0013 0.7792 0.4000 0.1940 0.0000 0.0000 12.7546 2.8764 12.0509 2.5789 1316.2361 312.5073 69.8734 4.5125 0.3600 0.0791 44.7083 0.9487 5.6419 1.6468 10.1378 0.0300 6.7442 1.1693 4.8253 1.7157 6.4691 4.5642 4.6830 4.8156 7.8353 2.7358 -0.7551 2.9605 3.3032 1.1785 3
20 6.1559 0.2683 0.4292 0.0062 8.9722 1.0288 97.3330 1.0703 98.5270 0.7454 89.0919 4.4202 114.0377 5.6578 88.0327 3.7638 89.7607 2.1110 7.2413 1.7748 0.4167 0.2308 0.0000 0.0000 6.1713 1.7867 6.1111 1.7159 712.5833 257.8134 74.4507 5.8188 0.1764 0.0549 42.2917 1.6544 4.9761 0.4564 8.1109 1.2902 5.3933 0.3402 2.7013 1.7991 4.2879 3.6772 2.7916 3.8070 5.0830 2.8560 -0.6349 2.6417 2.3969 1.4325 3
40 6.1333 0.1195 0.4392 0.0043 5.0926 0.9325 94.6637 1.6660 97.3410 1.1584 93.1589 0.6479 119.2433 0.8293 90.5746 0.7492 84.9920 1.2780 11.4620 1.7358 0.0600 0.0657 0.0000 0.0000 0.0787 0.0797 0.0185 0.0199 2.1852 1.9882 79.8933 84.3824 0.0023 0.0023 37.1528 0.7488 4.9248 1.9119 7.1829 1.1143 4.4644 1.3620 0.0320 1.2740 -0.0864 0.5983 -0.0165 0.2613 0.3143 0.0625 -0.2703 0.0392 0.1301 0.2936 3
}{\evalSthreeDMIG}
\pgfplotstableread{
period energy energy_ci epw epw_ci switch switch_ci acc acc_ci reqacc reqacc_ci pm pm_ci pmabs pmabs_ci gpu gpu_ci pack pack_ci cf cf_ci wait wait_ci waitp waitp_ci mig mig_ci inter inter_ci gbh gbh_ci copy copy_ci migvm migvm_ci calls calls_ci callmed callmed_ci callp95 callp95_ci wall wall_ci dacc dacc_ci dpm dpm_ci dgpu dgpu_ci dpack dpack_ci denergy denergy_ci depw depw_ci n
900 6.1910 0.1659 0.4297 0.0050 5.1574 1.2240 98.5453 1.6335 99.2963 0.7339 88.5943 3.9809 113.4007 5.0956 81.5604 2.4402 97.3043 0.2925 1.3043 0.1540 0.0000 0.0000 0.0000 0.0000 19.8889 1.3841 19.6343 1.2731 1085.2778 88.8426 35.3720 0.7762 0.5637 0.0425 8.0000 0.0000 0.0076 0.0017 0.0218 0.0039 0.0044 0.0004 3.9137 3.5825 4.8225 3.0872 9.9384 2.2449 12.6267 0.9795 -1.2148 2.0572 2.2830 1.4706 3
3600 6.2234 0.2858 0.4322 0.0042 3.9630 1.1043 98.0397 1.8410 99.0350 0.8220 90.1807 5.4341 115.4313 6.9556 82.3464 3.1840 96.3160 0.3120 2.1970 0.2441 0.0000 0.0000 0.0000 0.0000 9.9722 1.2720 9.8565 1.3104 530.2407 72.7969 34.4319 1.6017 0.2834 0.0381 2.0000 0.0000 0.0078 0.0005 0.0207 0.0060 0.0027 0.0007 3.4080 3.0965 3.1202 4.6108 9.0711 3.0351 11.6383 1.2726 -1.7384 3.0292 1.7109 1.2681 3
}{\evalSthreeTC}
\newcommand{\evalSfourFullAcc}{99.5}
\newcommand{\evalSfourFullPmabs}{111.4}

\newcommand{\evalSfourFullCalls}{44.7}

\newcommand{\evalSfourNoAdmAcc}{97.8}

\newcommand{\evalSfourNoConsAcc}{96.5}
\newcommand{\evalSfourNoConsPmabs}{119.5}

\newcommand{\evalSfourTimerPmabs}{114.1}

\newcommand{\evalSfourRandomPmabs}{110.0}

\newcommand{\evalSfourFFPmabs}{119.1}

\newcommand{\evalSfourMmaxFourPmabs}{118.8}

\newcommand{\evalSfourMmaxEightPmabs}{114.8}

\newcommand{\evalSfourMmaxThirtyTwoPmabs}{113.1}

\newcommand{\evalSfourMmaxThirtyTwoCallmed}{10.25}

\newcommand{\evalSfourLimitOnePmabs}{117.0}

\newcommand{\evalSfourLimitFivePmabs}{112.3}

\newcommand{\evalSfourLimitThirtyPmabs}{111.2}

\newcommand{\evalSfourThetaTwoPmabs}{111.7}

\newcommand{\evalSfourThetaFourPmabs}{112.3}

\newcommand{\evalSfourThetaFourCalls}{38.0}

\newcommand{\evalSfourRows}{%
			OrigaMIG & 3 & 99.5 & 111.4 & 6.16 & 12.8 & 44.7 \\ \hline
			without admission calls & 3 & 97.8 & 111.0 & 6.11 & 11.5 & 43.4 \\ \hline
			without consolidation calls & 3 & 96.5 & 119.5 & 6.20 & 0.9 & 1.1 \\ \hline
			timer instead of threshold & 3 & 97.5 & 114.1 & 6.16 & 6.2 & 8.8 \\ \hline
			random neighborhoods & 3 & 99.5 & 110.0 & 6.17 & 13.5 & 41.7 \\ \hline
			without calls (FF) & 3 & 94.6 & 119.1 & 6.12 & 0 & 0 \\ \hline
			\( M_{\max} = 4 \) & 3 & 95.3 & 118.8 & 6.16 & 2.0 & 38.6 \\ \hline
			\( M_{\max} = 8 \) & 3 & 97.0 & 114.8 & 6.14 & 6.4 & 41.4 \\ \hline
			\( M_{\max} = 32 \) & 3 & 98.5 & 113.1 & 6.19 & 11.6 & 41.8 \\ \hline
			time limit 1\,s & 3 & 96.4 & 117.0 & 6.17 & 3.8 & 39.6 \\ \hline
			time limit 5\,s & 3 & 98.2 & 112.3 & 6.15 & 9.6 & 42.9 \\ \hline
			time limit 30\,s & 3 & 98.9 & 111.2 & 6.14 & 12.0 & 44.0 \\ \hline
			\( \theta = 2 \) & 3 & 99.1 & 111.7 & 6.17 & 12.1 & 43.2 \\ \hline
			\( \theta = 4 \) & 3 & 98.7 & 112.3 & 6.18 & 10.7 & 38.0 \\ \hline}

\newcommand{\evalSfourScaleCallmedOneTwoEight}{4.26}

\newcommand{\evalSfourScaleCallsOneTwoEight}{44.7}

\newcommand{\evalSfourScaleDpmOneTwoEight}{2.8}

\newcommand{\evalSfourScaleDpmFiveOneTwo}{2.1}

\newcommand{\evalSfourScaleCallmedOneZeroTwoFour}{9.97}

\newcommand{\evalSfourScaleCallsOneZeroTwoFour}{207.8}

\newcommand{\evalSfourScaleDpmOneZeroTwoFour}{1.0}

\newcommand{\evalSfourScaleCallsTwoZeroFourEight}{284.9}

\newcommand{\evalSfourScaleDpmTwoZeroFourEight}{0.2}

\newcommand{\evalSfourScaleCallsFourZeroNineSix}{334.5}

\pgfplotstableread{
pms callmed callmed_ci callp95 callp95_ci calls calls_ci acc acc_ci mig mig_ci wait wait_ci dpm dpm_ci dacc dacc_ci n
128 4.2648 0.0000 10.1055 0.0000 44.6667 0.0000 99.5970 0.0000 11.7083 0.0000 1.1400 0.0000 2.7672 0.0000 7.5050 0.0000 1
512 6.4573 0.0000 10.1554 0.0000 129.7500 0.0000 99.4620 0.0000 9.6667 0.0000 0.0000 0.0000 2.0825 0.0000 0.2110 0.0000 1
1024 9.9697 0.0000 10.2695 0.0000 207.7500 0.0000 100.0000 0.0000 15.4583 0.0000 0.0000 0.0000 0.9785 0.0000 0.0000 0.0000 1
2048 9.5036 0.0000 10.2795 0.0000 284.9167 0.0000 100.0000 0.0000 18.6250 0.0000 0.0000 0.0000 0.2321 0.0000 0.0000 0.0000 1
4096 8.1779 0.0000 10.2263 0.0000 334.4583 0.0000 100.0000 0.0000 8.2083 0.0000 0.0000 0.0000 0.4459 0.0000 0.0000 0.0000 1
}{\evalSfourScale}
\newcommand{\evalStwoDemandLognormal}{160}
\newcommand{\evalSoneDemandZeropSixFive}{158}

\usepackage[ruled,vlined,linesnumbered]{algorithm2e}
\SetAlFnt{\small}
\SetAlCapFnt{\small}
\SetAlCapNameFnt{\small}
\SetFuncSty{textsc}
\DontPrintSemicolon

\newcommand{\CF}{\mathit{CF}}

\newcommand{\migsw}{1.05}   
\newcommand{\migrp}{0.50}   
\newcommand{\migby}{-0.30}  
\newcommand{\miglbl}{\sffamily\scriptsize}

\newcommand{\migfree}[1]{%
	\fill[black!12] ({#1*\migsw},0) rectangle ({(#1+1)*\migsw},0.42);}

\newcommand{\migbody}{%
	\foreach \i in {0,...,7} {%
		\draw ({\i*\migsw},0) rectangle ({(\i+1)*\migsw},0.42);
		\node[font=\miglbl] at ({(\i+0.5)*\migsw},0.21) {\i};}
	\draw (0,{\migby-0.50}) rectangle ({8*\migsw},\migby);
	\node[font=\miglbl\bfseries] at ({4*\migsw},{\migby-0.25}) {NVIDIA A100 (MIG-Enabled)};}

\newcommand{\miggi}[5][]{%
	\draw[#1] ({#3*\migsw},{0.50+#2*\migrp}) rectangle ({(#3+#4)*\migsw},{0.92+#2*\migrp});
	\node[font=\miglbl] at ({(#3+#4/2)*\migsw},{0.71+#2*\migrp}) {#5};}

\newcommand{\migx}[2]{%
	\node[font=\miglbl] at ({#2*\migsw},{0.71+#1*\migrp}) {$\times$};}

\begin{document}
	\begin{frontmatter}
		
		\title{OrigaMIG: MIG-Aware VM Placement with a Neighborhood-Restricted BILP and Live Migration}
		
		\author[ind]{Ahmad Siavashi}
		\ead{a.siavashi@outlook.com}
		
		\author[aff]{Mahmoud Momtazpour}
		\ead{momtazpour@aut.ac.ir}
		
		\address[ind]{Independent Researcher}
		\address[aff]{Department of Computer Engineering, Amirkabir University of Technology}
		
		\journal{Future Generation Computer Systems}
		
			\begin{abstract}
	The extensive use of GPUs in cloud computing, accelerated by the spread of large language model (LLM) services, and the growing need for multitenancy have driven the development of innovative solutions for efficient GPU resource management. Multi-Instance GPU (MIG) technology from NVIDIA enables shared GPU usage in cloud data centers by providing isolated instances, which are offered as MIG-backed virtual GPUs (vGPUs). However, MIG placement rules often lead to fragmentation and suboptimal resource allocation. In this work, we formally model the MIG-aware virtual machine (VM) placement as a binary integer linear programming (BILP) problem aimed at maximizing request acceptance, consolidating resources, and reducing migration overhead. Building upon this formulation, we propose OrigaMIG, a MIG-aware placement optimizer. OrigaMIG places each arriving request at once with a lightweight greedy rule and calls the solver only when a request cannot be placed or the GPUs of a class become fragmented. Each call solves the BILP on a small neighborhood of physical machines (PMs), keeps the rest of the data center fixed, and executes the solution as live migrations. We compare OrigaMIG with the default placement and with two state-of-the-art MIG-aware policies in simulation, across six loads, eleven variants of the workload and the hardware, and data centers of up to 4096 PMs. On 256 A100 and A30 GPUs in 128 PMs, OrigaMIG keeps fewer PMs active than the stronger policy in \evalSoneVsTCWins{} of \evalSoneVsTCRuns{} paired runs while migrating \evalSoneVsTCDmigMin\% to \evalSoneVsTCDmigMax\% fewer VMs, and it uses the least energy per admitted GPU-hour at every load. Against the default placement, it keeps up to \evalSoneGainDMIGDpmZeropFive\% fewer PMs active and admits up to \evalSoneGainDMIGDaccOne{} percentage points more of the requested GPU memory. On small data centers, it stays within 4.1\% of the optimal number of active PMs.
		\end{abstract}
		
		
	\end{frontmatter}
	
	\section{Introduction}
	\label{s:introduction}
	
	The rapid expansion of machine learning, data analytics, and high-performance computing (HPC) workloads has positioned GPUs at the heart of cloud computing services, a role amplified by the spread of self-hosted LLM services \cite{hou2025unveiling}. While GPUs offer exceptional parallel processing capabilities, they also introduce challenges in scheduling and resource sharing. Traditional scheduling methods, designed for standalone, monolithic GPUs, struggle to efficiently allocate resources when multiple tenants need to share a single GPU \cite{siavashi2023gvmp,chung_fine-grained_2025,amaral2017,zhu2021,li2022miso}.
	
	NVIDIA’s MIG technology, available on Ampere and Hopper GPUs \cite{nvidia_mig_doc}, partially addresses this challenge by partitioning a single GPU into multiple isolated and heterogeneous instances, each with dedicated memory and compute slices. This approach enhances performance isolation, enabling multiple workloads to run on the same GPU. However, MIG enforces rigid rules for slice alignment and profile placement, which can lead to fragmentation, where memory slices become unusable or inefficiently allocated, ultimately reducing overall resource efficiency \cite{lee2024parvagpu}.
	
	The NVIDIA A100 40GB, a MIG-enabled GPU, comprises 7 compute slices and 8 memory slices, each providing 5 GB of memory for a total capacity of 40 GB. It supports six distinct MIG GPU instance (GI) profiles, as shown in Table \ref{tab:a100-gis}, with each profile combining specific numbers of compute slices and memory slices. These profiles follow a naming convention, where the first digit is the number of compute slices and the second number is the memory in GB. While enabling MIG mode is a necessary first step, creating GPU instances and their corresponding compute instances (CIs) is also required. However, this paper focuses exclusively on GIs, leaving details about CIs outside its scope. A VM receives a GI as a MIG-backed vGPU \cite{nvidia_grid_doc}, with one vGPU per GI; the required profile is set by the workload the VM hosts, not by the scheduler. NVIDIA vGPU software supports live migration of such a VM, and requires that the destination GPU hosts a GI of the same profile.
	
	\begin{table}[h]
		\centering
		\scriptsize
		\caption{GPU instance profiles on A100 40GB}
		\begin{tabular}{|c|c|c|c|c|}
			\hline
			\textbf{Profile} & \textbf{Compute} & \textbf{Memory} & \textbf{Instances} & \textbf{Start Slices} \\ \hline
			1g.5gb            & 1/7                         & 1/8                                  & 7                                     & 0, 1, 2, 3, 4, 5, 6 \\ \hline
			1g.10gb           & 1/7                         & 2/8                                  & 4                                     & 0, 2, 4, 6 \\ \hline
			2g.10gb           & 2/7                         & 2/8                                  & 3                                     & 0, 2, 4 \\ \hline
			3g.20gb           & 3/7                         & 4/8                                  & 2                                     & 0, 4 \\ \hline
			4g.20gb           & 4/7                         & 4/8                                  & 1                                     & 0 \\ \hline
			7g.40gb           & 7/7                         & 8/8                                  & 1                                     & 0 \\ \hline
		\end{tabular}
		\label{tab:a100-gis}
	\end{table}

	Specific rules determine the allocation of GIs on MIG-enabled A100 GPUs, as detailed in NVIDIA's documentation \cite{nvidia_mig_doc}. This work adopts a perspective focused on memory slices rather than compute slices for greater intuitiveness. The start slices in Table~\ref{tab:a100-gis} already account for the compute slices, since the number of instances of a profile is the smaller of the bounds set by the memory slices and the compute slices of the GPU. No arrangement of profiles on the memory slices therefore exceeds the compute slices. These allocation principles are followed in other MIG-enabled GPUs.
	
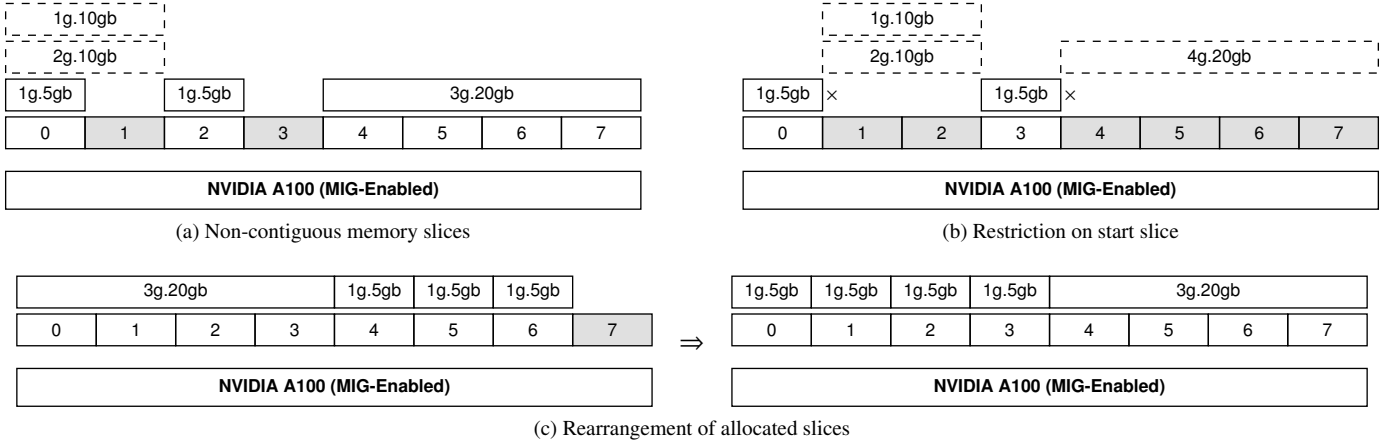
\begin{figure*}[t]
		\centering
		\subfloat[Non-contiguous memory slices]{
			
			\begin{tikzpicture}[line width=0.5pt]
				\migfree{1}\migfree{3}
				\migbody
				\miggi{0}{0}{1}{1g.5gb}
				\miggi{0}{2}{1}{1g.5gb}
				\miggi{0}{4}{4}{3g.20gb}
				\miggi[dashed]{1}{0}{2}{2g.10gb}
				\miggi[dashed]{2}{0}{2}{1g.10gb}
			\end{tikzpicture}
			\label{fig:frag1}
		}
		\hfill
		\subfloat[Restriction on start slice]{
			
			\begin{tikzpicture}[line width=0.5pt]
				\foreach \z in {1,2,4,5,6,7} {\migfree{\z}}
				\migbody
				\miggi{0}{0}{1}{1g.5gb}
				\miggi{0}{3}{1}{1g.5gb}
				\migx{0}{1.12}\migx{0}{4.12}
				\miggi[dashed]{1}{1}{2}{2g.10gb}
				\miggi[dashed]{1}{4}{4}{4g.20gb}
				\miggi[dashed]{2}{1}{2}{1g.10gb}
			\end{tikzpicture}
			\label{fig:frag2}
		}\
		\subfloat[Rearrangement of allocated slices]{
			
			\begin{tikzpicture}[line width=0.5pt]
				\migfree{7}
				\migbody
				\miggi{0}{0}{4}{3g.20gb}
				\miggi{0}{4}{1}{1g.5gb}
				\miggi{0}{5}{1}{1g.5gb}
				\miggi{0}{6}{1}{1g.5gb}
			\end{tikzpicture}%
			\hspace{0.35cm}\raisebox{0.75cm}{$\Rightarrow$}\hspace{0.35cm}%
			\begin{tikzpicture}[line width=0.5pt]
				\migbody
				\foreach \z in {0,...,3} {\miggi{0}{\z}{1}{1g.5gb}}
				\miggi{0}{4}{4}{3g.20gb}
			\end{tikzpicture}
			\label{fig:frag3}
		}
		\caption{MIG fragmentation scenarios}
		\label{fig:a100-frag}
	\end{figure*}

	The rules for placing MIG profiles present challenges in the allocation of GIs. 	GIs are created and destroyed in response to accepted and completed requests, from short batch jobs to the long-lived tenants of self-hosted LLM serving, which can lead to fragmentation. Fig.~\ref{fig:a100-frag}(a) depicts a common fragmentation scenario where, due to non-contiguous memory slices, it is impossible to allocate either 1g.10gb or 2g.10gb GIs. In MIG, however, having contiguous free memory slices does not always ensure GI allocation. Fig.~\ref{fig:a100-frag}(b) depicts a situation where the displayed MIG profiles cannot be assigned to the specified contiguous slices, owing to the lack of availability of their necessary start slices. Furthermore, 
	Fig.~\ref{fig:a100-frag}(c) illustrates a scenario where compacting the allocated slices on the GPU allows for the assignment of an additional 1g.5gb profile. These illustrations show traditional external fragmentation and the placement fragmentation specific to MIG, where contiguous slices remain free but their necessary start slices are unavailable.
	
	In large-scale cloud environments, fragmentation worsens as MIG partitions are continuously created and removed to accommodate shifting user demands. GPUs account for most of the cost of these machines, and their placement rules leave capacity unusable where CPU and RAM would not. A naive or short-sighted allocation strategy can quickly lock GPUs into a suboptimal state, limiting the acceptance of high-priority requests and increasing the number of active GPUs. To our knowledge, no existing placement strategy accounts for MIG’s strict placement constraints when placing VMs across a heterogeneous, multi-host data center.
	
	This article makes the following contributions.
	\begin{itemize}
		\item An analysis of the default GI placement policy of MIG (Section~\ref{s:default-placement-policy}). We analyze the configuration flexibility (\( \CF \)) of a GPU, its number of free start slices summed over the GI profiles, show that the default policy puts each GI where it maximizes \( \CF \), and find that 67\% of the 723 configurations of an A100 arrange their GIs suboptimally. As a count of free start slices, \( \CF \) enters the objective of the BILP as a linear term.
		\item A BILP model of MIG-aware VM placement across a heterogeneous data center with several GPUs per PM (Section~\ref{s:formulation}). It models the start slices of each GI profile and the CPU and RAM of the PMs, and its objective weighs the admitted requests, the active PMs and GPUs, and the migrations of running VMs.
		\item OrigaMIG, an online optimizer built on this model (Section~\ref{s:methodology}). It admits requests on a fast greedy path and rearranges GIs on a slow path. A request that the greedy rule cannot place triggers an admission call, and a fragmented class of GPUs triggers a consolidation call. Each call solves the BILP on a neighborhood of at most 16 PMs with the rest of the data center fixed, and executes the solution as an ordered sequence of live migrations of MIG-backed vGPUs \cite{nvidia_grid_doc}. OrigaMIG never removes a running VM.
		\item An evaluation in simulation against the default placement and two state-of-the-art MIG-aware policies, MFI \cite{zambianco2025online} and TC \cite{turkkan2024optimal}, adapted to VMs on PMs (Section~\ref{s:evaluation}). The default placement, FF, pairs first-fit host selection with the default GI placement of Section~\ref{s:default-placement-policy}. The evaluation covers loads from 0.5 to 1.2 on 128 PMs with A100 and A30 GPUs, eleven variants of the workload and the hardware, including a production trace, the cost of migrations, an ablation of OrigaMIG, data centers of up to 4096 PMs, and the energy of the PMs and GPUs.
		\item A comparison of OrigaMIG with the optimal placement on data centers of 8 to 32 PMs, and a measurement of how the time to solve the BILP over a whole data center grows (Sections~\ref{s:optgap} and \ref{s:bilp-scalability}).
	\end{itemize}
	
	On 128 PMs, about the size of one partition of a data center such as a row or a pod, OrigaMIG keeps fewer PMs active than TC, the stronger baseline, in \evalSoneVsTCWins{} of \evalSoneVsTCRuns{} paired runs over six loads, while migrating \evalSoneVsTCDmigMin\% to \evalSoneVsTCDmigMax\% fewer VMs. Against the default placement (FF), it keeps up to \evalSoneGainDMIGDpmZeropFive\% fewer PMs active, against \evalSoneGainTCDpmZeropFive\% for TC, admits up to \evalSoneGainDMIGDaccOne{} percentage points more of the requested GPU memory, and uses the least energy per admitted GPU-hour at every load. Across eleven variants of the workload and the hardware, OrigaMIG keeps fewer PMs active than TC in seven, about as many in one, and more in the three with the largest VMs relative to their GPUs. One instance of OrigaMIG serves up to about 1024 PMs, so a larger data center runs one instance per partition of its PMs. On data centers of 8 to 32 PMs, OrigaMIG stays within 4.1\% of the optimal number of active PMs, with calls of about one second or less.

	The remainder of this paper is structured as follows. Section \ref{s:related-work} reviews related work, while Section \ref{s:default-placement-policy} describes the default GI placement policy. Section \ref{s:formulation} formalizes the problem as a BILP model, and Section \ref{s:methodology} presents our proposed optimizer, OrigaMIG, which rearranges GIs by solving the BILP on small neighborhoods of PMs. Section~\ref{s:evaluation} provides an evaluation of OrigaMIG, Section~\ref{s:discussion} discusses its use and its limits, and Section~\ref{s:conclusion} concludes with future directions.
	
	\section{Related Work}
	\label{s:related-work}
	
	The problem of VM placement in cloud data centers is well studied \cite{alashaikh2021,dias2021,imran2022,saidi2023task,lin2024energy,alahmad2024multiple,xu2024meta,regaieg2021multi,talebian2020optimizing,masdari2020bio,kong2022nature,sudhakar2021survey,saxena2021secure,caviglione2021deep,torre2020dynamic,wei2020exact,shaw2019energy}. However, GPU virtualization introduces unique constraints that do not arise with CPUs, leading to additional challenges \cite{hong2017gpu}. As a result, GPU-enabled VM placement becomes more complex and challenging.
	
	\emph{GPU-Sharing VM Placement (GRID Era).} NVIDIA introduced GRID \cite{nvidia_grid_doc}, whose early generations required co-resident vGPU partitions to be homogeneous. In \cite{siavashi2019gpucloudsim}, NVIDIA GRID is examined for GPU-enabled VM placement, highlighting inefficiencies in the first-fit strategy as VM numbers grow. The proposed first-fit increasing algorithm improves acceptance rate (59\%), makespan (25\%), and energy consumption (21\%). In \cite{kulkarni2021gpu}, NVIDIA GRID is further analyzed for GPU-aware VM placement, enhancing energy efficiency, reducing host search space, and shortening makespan while tackling GPU resource management challenges in cloud applications. An integer linear programming model for GPU-enabled VM placement using NVIDIA GRID GPU virtualization is proposed in \cite{garg2019virtual}, aiming to minimize physical GPU usage in the cloud. Two heuristics match its performance with faster execution. A metaheuristic approach using dense neural networks (DNN) is proposed in \cite{sivaraman2019tecn}. The authors build a simulator for GPU-enabled VM placement with NVIDIA GRID, implement six heuristics, and train DNNs on various configurations. They also explore combining multiple DNNs. Using a custom metric based on GPU utilization and job waiting time, their method outperforms heuristics in 76\% of test cases. In \cite{sivaraman2018task}, authors explore a GPU-enabled VM placement problem for NVIDIA GRID, optimizing performance using a cost function based on GPU utilization, fully executed VMs, and GPU execution time. Various VM placement policies are analyzed, with one achieving the best results.

	\emph{MIG Partitioning and Co-location.} Later, NVIDIA introduced MIG \cite{nvidia_mig_doc}, which enabled both homogeneous and heterogeneous instances; however, it enforces a strict set of placement rules, making placement more complex than the earlier GRID solution. To optimize MIG partitions for serving DNNs, Tan et al.\ \cite{tan2021serving} introduce the reconfigurable machine scheduling problem, a new NP-hard abstraction. They propose MIG-serving, a pipeline combining greedy heuristics, genetic algorithm, and Monte Carlo tree search for cost-efficient deployments. MIG-serving includes an optimizer for deployment generation and a controller for seamless application. Experiments on a 24-GPU A100 cluster show up to a 40\% GPU reduction over baseline static allocations with minimal service disruptions. In \cite{arima2022optimizing}, the authors integrate GPU partitioning with power-aware scheduling to optimize resource allocation in CPU-GPU heterogeneous systems under power constraints. Leveraging MIG for fine-grained workload co-location, their approach employs scalability and interference models to improve efficiency and achieve near-optimal configurations across diverse workloads.
	
	In \cite{li2022miso}, authors highlight the rapid advancements in GPU technology, enhancing HPC and AI/ML research but also leading to inefficient resource utilization. To address this, they propose MISO, a technique leveraging MIG on NVIDIA datacenter GPUs (e.g., A100, H100) for dynamic GPU resource partitioning. MISO utilizes the lightweight Multi-Process Service (MPS) \cite{nvidia_mps_doc} to predict optimal MIG partitioning without incurring exploration overhead. This approach improves GPU efficiency, achieving 49\% and 16\% lower average job completion times (JCT) compared to unpartitioned and optimal static GPU partitioning, respectively. In \cite{saroliya2023hierarchical}, authors integrate MPS and MIG hierarchically, using reinforcement learning to jointly optimize partitioning and job co-scheduling. Experimental results demonstrate up to 1.87$\times$ throughput improvement over time-sharing scheduling.
	
	In \cite{zhang2024}, authors present MIGER, a scheduling framework that combines MIG's hardware partitions with MPS concurrency to improve job turnaround in GPU clusters. The authors model the MIG-enabled placement of jobs as an NP-hard optimization and design both exact and heuristic algorithms to allocate GPU instances. Through real cluster experiments, MIGER achieved 36\% and 46.6\% faster job completion times than state-of-the-art MIG-based and MPS-based solutions, respectively. Similarly, in \cite{lee2024parvagpu}, the authors propose ParvaGPU, a GPU space-sharing framework that integrates MIG and MPS to optimize DNN inference in cloud environments. It mitigates GPU underutilization with the Optimal Triplet Decision algorithm, which selects MPS-activated MIG instances for maximum throughput. The Demand Matching algorithm optimally assigns instances to meet high request rates. To reduce GPU non-contiguous space inefficiency, Segment Relocation redistributes instances, while Allocation Optimization minimizes fragmentation by resizing partitions. Evaluations on A100 GPUs show that ParvaGPU ensures service-level objective (SLO) compliance while reducing GPU usage.
	
	\emph{Placement, Fragmentation, and Migration.} Tang et al.\ \cite{tangSTCCC25} tackle GPU sharing from an I/O perspective by accounting for PCIe data-transfer bottlenecks in MIG-enabled environments. They introduce a PCIe-bandwidth-aware scheduler that places MIG partitions so as to avoid PCIe-bandwidth contention, thereby preventing performance degradation when multiple partitions contend for host-to-GPU links. In \cite{weng_beware_2023}, in the fractional-sharing setting, authors identify GPU fragmentation as a key inefficiency in large ML clusters, where traditional bin packing fails to address partial GPU allocations. They propose Fragmentation Gradient Descent (FGD), which quantifies fragmentation and schedules tasks along its steepest descent to optimize GPU use. Evaluated on a 6{,}200-GPU cluster, FGD reduces unallocated GPUs by up to 49\%, reclaiming 290 more GPUs than existing schedulers.
	
	Saraha et al.\ \cite{saraha2025managing} dynamically fuse and split MIG partitions at runtime to improve throughput and energy efficiency. Turning to fragmentation, Turkkan et al.\ \cite{turkkan2024optimal} formulate MIG placement and migration as a mixed-integer program for large language model (LLM) inference workloads. Two online fragmentation-aware MIG schedulers have also been proposed \cite{zambianco2025online,ting2025online}: the former minimizes a fragmentation metric through single-pass placement, while the latter combines conditional load balancing with dynamic repartitioning and migration under a makespan objective.

	Villarrubia et al.\ \cite{villarrubia2025leveraging,solving2025rlmig} study offline moldable-task MIG scheduling, first with an approximation-bound heuristic and subsequently with deep reinforcement learning. In the fractional GPU-sharing setting, Lettich et al.\ \cite{lettich2024power} extend FGD with power awareness, while Wu et al.\ \cite{shen2025defragmentation} apply deep reinforcement learning to defragment shared GPU clusters. Finally, Flex-MIG \cite{kim2025flexmig} relaxes MIG's one-to-one execution model using host-side memory collectives, a complementary direction that future placement schedulers may exploit.

	\begin{table*}[t]
	\centering
	\caption{MIG-aware placement and scheduling approaches. The parenthesis in Placed unit gives the form in which the unit runs, and Migration lists the entities moved or reconfigured at runtime.}
	\label{tab:related-work}
	\scriptsize
	\setlength{\tabcolsep}{3pt}
	\renewcommand{\arraystretch}{1.15}
	\newcolumntype{R}{>{\raggedright\arraybackslash}X}
	\begin{tabularx}{\textwidth}{|l|c|>{\hsize=1.45\hsize}R|c|c|>{\hsize=0.95\hsize}R|c|>{\hsize=0.5\hsize}R|>{\hsize=1.1\hsize}R|}
	\hline
	\textbf{Work} & \textbf{Sharing} & \textbf{Placed unit} & \textbf{Profile chosen by} & \textbf{GIs/unit} & \textbf{Infrastructure} & \textbf{Planning} & \textbf{Migration} & \textbf{Objective} \\ \hline
	MIG-serving \cite{tan2021serving}            & MIG     & DNN services (pods)          & Scheduler & Several & Homog.\ cluster          & Periodic & Pods + GIs & GPUs used \\ \hline
	MIGER \cite{zhang2024}                       & MIG+MPS & DL jobs (MPS processes) & Scheduler & One     & Homog.\ cluster          & Online   & GIs                    & JCT under QoS \\ \hline
	ParvaGPU \cite{lee2024parvagpu}              & MIG+MPS & DNN inference services (MPS processes) & Scheduler & Several & Homog.\ cluster          & Offline  & GIs                    & GPU usage under SLO \\ \hline
	Turkkan et al.\ \cite{turkkan2024optimal}    & MIG     & LLM replicas (pods) & Request   & One     & Homog.\ cluster          & Periodic & Pods + GIs & Active GPUs, migrations \\ \hline
	Zambianco et al.\ \cite{zambianco2025online} & MIG     & Generic requests (unspecified) & Request   & One     & Homog.\ GPU pool         & Online   & None                   & Fragmentation \\ \hline
	Ting et al.\ \cite{ting2025online}           & MIG     & LLM inference jobs (unspecified) & Request   & One     & Single node              & Online   & Jobs + GIs & Makespan \\ \hline
	\textbf{OrigaMIG}                                & MIG     & VMs                          & Request   & Several & Hetero.\ data center & Online   & VMs + GIs              & Acceptance, consolidation, migrations \\ \hline
	\end{tabularx}
	\end{table*}

	Table~\ref{tab:related-work} summarizes how OrigaMIG differs from the works that account for the MIG placement constraints. Unlike the surveyed approaches, which schedule jobs or inference requests, OrigaMIG places VMs that request whole GPU instances, while two of them further sub-divide each GPU through MPS. The three serving systems choose the MIG profile of each service from offline profiles, whereas OrigaMIG, like the fragmentation-aware schedulers, takes the profile as part of the request. Among these works, OrigaMIG is also the only one to model heterogeneous, multi-host infrastructure, capturing host CPU and memory across GPUs that may differ in type and generation, where the others assume a single node or a homogeneous cluster. Finally, OrigaMIG migrates VMs both within and across GPUs and hosts and it jointly optimizes acceptance rate, consolidation, and migration overhead, treating the number of migrations as an explicit objective rather than a by-product of reconfiguration.
	
	\section{Default GI Placement Policy}
	\label{s:default-placement-policy}
	
	In this section, we examine how the default GI placement policy affects a GPU’s ability to host future profiles. We use the configuration flexibility (\( \CF \)) metric \cite{siavashi2025multi} to quantify this flexibility, analyze how default placements influence it, and highlight common inefficiencies. These insights highlight the challenges involved in designing VM placement strategies that align with NVIDIA’s fixed MIG management rules.
	
	A configuration \( G \) is the set of GI placements on a GPU, each represented as \( \text{Start}{:}\text{Size} \) in memory slices, so placing a profile \( p \) of size \( g_p \) at start slice \( z \) adds the placement \( z{:}g_p \) to it. Let \( S(F, p) \) denote the set of permissible start slices of \( p \) in a set \( F \) of free memory slices.
	
	\begin{align}
		\begin{split}
			\label{eq:s}
			S(F, p) = \{ z \in g_p \mathbb{Z}_{\geq 0} \mid z \leq s_p, \, \{ z, \dots, z + g_p - 1 \} \subseteq F \}
		\end{split}
	\end{align}
	
	A start slice is a multiple of \( g_p \), and the \( g_p \) slices the GI occupies are all free. The limit \( s_p \) is the last start slice permitted by the compute slices of the GPU. Given a set \( \mathcal{P} \) of all supported profiles (see Table~\ref{tab:a100-gis}) and the slices \( F \) left free by \( G \), \( \CF(G) \) is then defined as the total number of possible placements for all profiles within the current GPU configuration as follows:
	
	\begin{align}
		\label{eq:cc}
		\CF(G) = \sum_{p \in \mathcal{P}} |S(F, p)|
	\end{align}	
	
	For example, the configuration \( G = \{0{:}1, 3{:}1\} \) in \text{Fig.~\ref{fig:a100-frag}(b)} leaves \( F = \{1, 2, 4, 5, 6, 7\} \) free and has a \( \CF(G) = 9 \) because it can accommodate \( 5 \times 1g.5\text{gb} \), \( 2 \times 1g.10\text{gb} \), \( 1 \times 2g.10\text{gb} \), and \( 1 \times 3g.20\text{gb} \) profiles. The default GI placement policy observed in our experiments with NVIDIA driver R530 positions a profile within the given configuration to maximize the \( \CF \) value. For a given GI profile \( p_{req} \), the policy selects the start slice \( z^{*} \) that maximizes \( \CF \), and the lowest such slice when several attain the maximum, as represented mathematically below:
	
	\begin{align}
		\label{eq:max-cc}
		z^{*} = \min \left\{ \underset{z \in S(F, p_{req})}{\text{arg max}} \CF(G \cup \{ z{:}g_{p_{req}} \}) \right\}
	\end{align}
	The profile cannot be placed when \( S(F, p_{req}) \) is empty.
	
	Let us now examine all possible MIG memory configurations for a single A100 GPU, both exhaustively and under the default GI placement policy, to highlight how GI placement affects resource efficiency. Using the $\CF$ metric, we show that default allocations frequently result in suboptimal configurations. We employ depth-first search to develop a tree that consists of all possible memory configurations for a single A100 GPU, starting from a state where the GPU is not in use. We generate a new child node for each configuration by adding a new GI. This addition of GIs continues until the GPU cannot accommodate any more, resulting in 78 terminal nodes where no further GIs can be added. The finalized tree encompasses 723 unique configurations. In practical scenarios, the addition or removal of GIs shifts the configuration of the GPU to one of these identified configurations.
	
	Configurations may have the same GIs but show different $\CF$ values based on how the GIs are arranged in their slices. Defining an optimal configuration as one that attains the highest $\CF$ value among all arrangements with the same set of GIs, it is found that 67\% of the 723 configurations, or 482 in total, are in suboptimal arrangements. In multi-GPU environments, the probability of encountering different GI profiles becomes more crucial. 
	
	The default GI placement policy, which sequentially assigns GIs, results in 179 configurations, representing 25\% of the total 723 possible configurations in the space, when GIs are only added. This sequential assignment approach can lead to suboptimal GPU states. Take, for instance, the arrival of two 1g.5gb GI profiles. Ideally, in an empty GPU, these GIs should occupy memory slices 4 and 5. However, treating GIs independently may lead to them being placed in slices 4 and 6, which is not the most efficient allocation. Approximately 33\%, equating to 59 out of the 179 allocations made by the default policy, fall short of being optimal. Moreover, since the policy does not reposition GIs post-allocation, the departure and subsequent arrival of GIs shift the configuration to any of the 723, and one that was initially optimal can become suboptimal.
	
\begin{figure}[h]
		\centering
		\subfloat[Configuration A]{
			
			\begin{tikzpicture}[line width=0.5pt]
				\foreach \z in {0,1,2,3,5,7} {\migfree{\z}}
				\migbody
				\miggi{0}{4}{1}{1g.5gb}
				\miggi{0}{6}{1}{1g.5gb}
			\end{tikzpicture}
			\label{fig:config-a}
		}\
		\subfloat[Configuration B]{
			
			\begin{tikzpicture}[line width=0.5pt]
				\foreach \z in {2,...,7} {\migfree{\z}}
				\migbody
				\miggi{0}{0}{1}{1g.5gb}
				\miggi{0}{1}{1}{1g.5gb}
			\end{tikzpicture}
			\label{fig:config-b}
		}
		\caption{Configurations with identical $\CF$ but different per-profile capacity}
		\label{fig:a100-alt-configs}
	\end{figure}
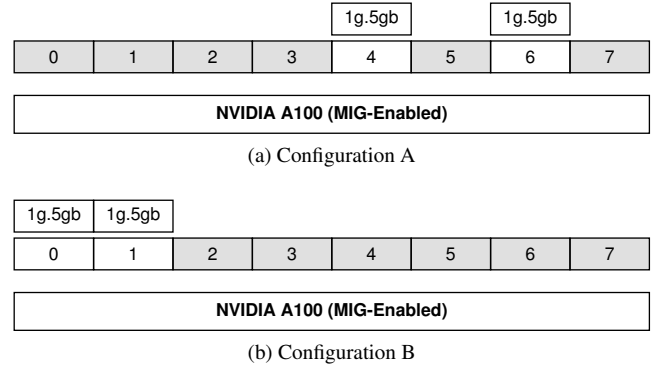

	The $\CF$ metric effectively indicates a configuration's capacity to accommodate new GIs, but it simplifies the optimality calculation by treating different profiles as equivalent. This approach neglects real-world conditions where certain GIs, such as 1g.10gb, might be more likely to occur. Contrary to what might be assumed, sufficient space for a larger profile does not guarantee the fit of smaller ones due to each profile's specific slice fitting requirements. Fig.~\ref{fig:a100-alt-configs} contrasts two GPU configurations with identical $\CF$ but different per-profile capacity. As highlighted in Table \ref{tab:alt-configs}, despite possessing the same $\CF$, Configuration B accommodates more 1g.10gb profiles, albeit at the expense of one fewer 4g.20gb profile.
	
	\begin{table}[h]
		\centering
		\scriptsize
		\caption{Per-profile capacity of the two configurations in Fig.~\ref{fig:a100-alt-configs}}
		\begin{tabular}{l|c|c}
			\textbf{Profile} & \textbf{Configuration A} & \textbf{Configuration B} \\
			\hline
			1g.5gb  & 5 & 5 \\
			1g.10gb & 2 & 3 \\
			2g.10gb & 2 & 2 \\
			3g.20gb & 1 & 1 \\
			4g.20gb & 1 & 0 \\
			7g.40gb & 0 & 0 \\
			\hline
			$\CF$ & 11 & 11 \\
			\hline
		\end{tabular}
		\label{tab:alt-configs}
	\end{table}
	
	In a single GPU with 723 unique configurations, 59\% or 428 configurations can support a specific profile more effectively than alternative configurations that have the same profiles arranged in different slices, despite these alternatives having the same or lower $\CF$ value. With two GPUs, there are $\binom{723 + 2 - 1}{2} = 261,726$ distinct configurations and in 79\%, or 205,575 cases, alternative configurations exist that can accommodate at least one type of profile more effectively, even if they have the same or lower $\CF$ compared to the original arrangement. With the addition of more GPUs, the complexity of the configuration space grows significantly, making a thorough examination increasingly difficult, or perhaps impractical.
	
	These analyses show that suboptimal configurations are common across the space and among the allocations the default policy makes. The default policy assigns one GI at a time on a single GPU and does not reposition it once allocated. This work therefore proposes a VM placement policy tailored for MIG-enabled cloud data centers, in which migration emerges as a necessary tool to achieve better allocations and mitigate suboptimal configurations.
	
	\section{Problem Definition and Objectives}
	\label{s:formulation}
	
	We model the problem as a BILP \cite{siavashi2025multi} in which requests arrive online, each with requirements and a duration that are unknown a priori, and placement decisions may also migrate previously placed VMs. We adopt a discrete-time approach in which the requests arriving in each interval are evaluated and placed alongside the VMs already residing. We solve each interval without knowledge of the requests that follow, which makes the setting non-clairvoyant. Time indices are omitted to enhance readability, and primed variables denote previous-interval state. We assume that a GI is created on the PM before its VM starts or arrives there by migration, so the model places each GI rather than selecting among instances that already exist. The model supports heterogeneity in workloads and infrastructure configurations, closely mirroring real-world environments.
	
	Consider $N$ as the set of VMs and $M$ as the set of physical machines (PMs). We refer to the $i^{th}$ VM in $N$ as $\text{VM}_i$ and the $j^{th}$ PM in $M$ as $\text{PM}_j$. The collection of GPUs in $\text{PM}_j$ is represented by $K_j$. The variable $x_{ij}$ is used to indicate the allocation of $\text{VM}_i$ to $\text{PM}_j$. The set $V_i$ holds the GIs that $\text{VM}_i$ requests, and is empty for a VM without a GPU. Each GI $v \in V_i$ has its own profile $p_v$, which sets its size $g_v$ and its compatibility class $h_v$. Similarly, $y_{ivjkz}$ represents the assignment of GI $v$ of $\text{VM}_i$ to the start slice $z$ of the $k^{th}$ GPU of $\text{PM}_j$, and $\mathcal{S}_{ivjk}$ is the set of permissible start slices for that GI. The essential symbols are provided in Table \ref{tab:notations}.
	
	\begin{align}
		\begin{split}
			\label{eq:8}
			& \mathcal{S}_{ivjk} = \begin{cases}
				S(U_{jk}, p_v) & \text{if } h_v = H_{jk} \\
				\emptyset & \text{otherwise}
			\end{cases} \\
			& \qquad \forall i \in N, \forall v \in V_i, \forall j \in M, \forall k \in K_j
		\end{split}
	\end{align}
	
	Eq. \eqref{eq:8} applies Eq. \eqref{eq:s} to GI $v$ of $\text{VM}_i$ on the $k^{th}$ GPU of $\text{PM}_j$, with all of its memory slices $U_{jk}$ free, so $\mathcal{S}_{ivjk}$ does not depend on the GIs already placed. The set is empty on a GPU of a different compatibility class, where the classes $h_v$ and $H_{jk}$ are categorical. We write \( o_{jku} = \sum_{i \in N} \sum_{v \in V_i} \sum_{z \in \mathcal{S}_{ivjk},\, z \leq u < z + g_v} y_{ivjkz} \) for the sum over the GIs that would cover slice \( u \) of the \( k^{th} \) GPU of \( \text{PM}_j \), which Eq.~\eqref{eq:10} limits to one, so \( o_{jku} \) indicates whether the slice is occupied. Using these definitions, the problem is formulated as follows:
	
	\begin{align}
		\notag & \textbf{Maximize} \quad \sum_{i \in N} \sum_{j \in M} a_{i} x_{ij} - \sum_{j \in M} b_j \varphi_{j} - \sum_{j \in M} \sum_{k \in K_j} e_{jk} \gamma_{jk} \\
		\label{eq:1} & \qquad - \sum_{i \in N} \delta_{i} (m_i + \omega_i) + \sum_{j \in M} \sum_{k \in K_j} \sum_{p \in \mathcal{P}_{jk}} \sum_{z \in S(U_{jk}, p)} w_p f_{jkpz} \\
		\notag & \textbf{Subject to} \\
		\label{eq:4} & \sum_{i \in N} x_{ij} c_{i} \leq C_{j} \quad \forall j \in M \\
		\label{eq:5} & \sum_{i \in N} x_{ij} r_{i} \leq R_{j} \quad \forall j \in M \\
		\label{eq:6} & \sum_{j \in M} x_{ij}^{'} \leq \sum_{j \in M} x_{ij} \leq 1 \quad \forall i \in N \\
		\label{eq:7} & \sum_{k \in K_j} \sum_{z \in \mathcal{S}_{ivjk}} y_{ivjkz} = x_{ij} \quad \forall i \in N, \forall v \in V_i, \forall j \in M \\
		\label{eq:10} & o_{jku} \leq \gamma_{jk} \quad \forall j \in M, \forall k \in K_j, \forall u \in U_{jk} \\
		\notag & f_{jkpz} + o_{jku} \leq 1 \quad \forall u \in \{z, \dots, z + g_p - 1\} \\
		\label{eq:11} & \qquad \forall j \in M, \forall k \in K_j, \forall p \in \mathcal{P}_{jk}, \forall z \in S(U_{jk}, p) \\
		\notag & f_{jkpz} \leq \gamma_{jk} \quad \forall j \in M, \forall k \in K_j, \forall p \in \mathcal{P}_{jk}, \forall z \in S(U_{jk}, p) \\
		\label{eq:17} & x_{ij} \leq \varphi_j \quad \forall i \in N, \forall j \in M \\
		\label{eq:19} & \gamma_{jk} \leq \sum_{i \in N} \sum_{v \in V_i} \sum_{z \in \mathcal{S}_{ivjk}} y_{ivjkz} \quad \forall j \in M, \forall k \in K_j \\
		\label{eq:20} & m_i \geq \sum_{j \in M} (1 - x_{ij}^{'}) x_{ij} \quad \forall i \in N \\
		\label{eq:21} & \omega_i \geq \sum_{j \in M} \sum_{k \in K_j} \sum_{z \in \mathcal{S}_{ivjk}} (1 - y_{ivjkz}^{'}) y_{ivjkz} \quad \forall i \in N, \forall v \in V_i \\
		\label{eq:24} & x_{ij}, y_{ivjkz}, \varphi_j, \gamma_{jk}, m_i, \omega_i, f_{jkpz} \in \{0,1\}
	\end{align}
	
	The first term of Eq. \eqref{eq:1} is designed to optimize the request acceptance by maximizing the sum of weights $a_i$ for each VM allocated. The second term aims to consolidate resources by minimizing the number of active PMs and GPUs, using $b_j$ to weight each PM, $e_{jk}$ to weight each GPU, and $\gamma_{jk}$ to indicate active GPUs. Consolidation reduces fragmentation, which in turn allows the acceptance of the requests that follow. The third term seeks to minimize the migration of VMs and their GIs weighted by $\delta_i$, thus enhancing system stability and efficiency. The fourth term extends the consolidation to the arrangement of GIs within a GPU by rewarding each start slice left free for a profile \( p \) with \( w_p \), where \( \mathcal{P}_{jk} \) holds the profiles of the compatibility class of the GPU. With all \( w_p \) equal, the term sums the \( \CF \) of the active GPUs, and unequal \( w_p \) rank the profiles by their expected demand. The weights place the four terms on a common scale. For example, \( a_i \) larger than the \( w_p \) of the start slices its GIs occupy makes its acceptance preferable to the free windows.

	Constraints set by Eq. \eqref{eq:4} and Eq. \eqref{eq:5} ensure that the CPU and RAM allocations for each PM do not exceed its respective capacities $C_j$ and $R_j$. Eq. \eqref{eq:6} guarantees that each VM is allocated to at most one PM, and that a VM allocated in the previous interval stays allocated, so the model may migrate a running VM but never evicts it, whereas Eq. \eqref{eq:7} ensures that each accepted VM receives each of its GIs on its own PM, and a rejected VM none. Eq. \eqref{eq:10} is designed to prevent overlapping GIs on the same GPU by allowing at most one GI on each memory slice. It also requires $\gamma_{jk}$ to be set whenever the GPU hosts a GI. Eq.~\eqref{eq:11} lets \( f_{jkpz} \) be set only when no GI occupies any of the \( g_p \) slices from \( z \), so at the optimum the sum of \( f_{jkpz} \) over \( z \) equals \( |S(F, p)| \) of Eq.~\eqref{eq:s} for the slices \( F \) left free. The bound \( f_{jkpz} \leq \gamma_{jk} \) keeps the reward off GPUs that host no GI, so at the optimum the term is the sum of the free windows of the active GPUs only. Eq. \eqref{eq:17} and Eq.~\eqref{eq:19} are used to determine $\varphi_j$ and $\gamma_{jk}$ ensuring that VMs are not assigned to powered-off PMs. Eq. \eqref{eq:20} and Eq.~\eqref{eq:21} count, per VM, whether the VM (resp.\ any of its GIs) is placed on a PM (resp.\ a GPU and start slice) that it did not previously occupy, and a single inter-machine migration counts once. A GI is created before its VM starts, so a different slice is a different GI and the change is counted as a migration. Finally, Eq.~\eqref{eq:24} defines the domains of all variables.
	
	\begin{table}[h]
		\centering
		\scriptsize
		\caption{Table of notations}
		\begin{tabular}{c|l}
			\hline
			\textbf{Notation} & \textbf{Description} \\
			\hline
			\( N \) & Set of VMs \\
			\( M \) & Set of PMs \\
			VM\(_i\) & The \( i^{th} \) VM in the set \( N \) \\
			PM\(_j\) & The \( j^{th} \) PM in the set \( M \) \\
			\( K_j \) & Collection of GPUs in PM\(_j\) \\
			\( U_{jk} \) & Collection of memory slices in GPU \( k \) of PM\(_j\) \\
			\( x_{ij} \) & Indicates allocation of VM\(_i\) to PM\(_j\) \\
			\( x_{ij}^{'} \) & Represents previous allocation state of VM\(_i\) to PM\(_j\) \\
			\( y_{ivjkz} \) & Represents assignment of GI \( v \) of VM\(_i\) to start slice \( z \) of GPU \( k \) \\
			\( y_{ivjkz}^{'} \) & Previous assignment of GI \( v \) of VM\(_i\) to start slice \( z \) of GPU \( k \) \\
			\( a_i \) & Weight associated with VM\(_i\) in the objective function \\
			\( b_j \) & Weight associated with PM\(_j\) in the objective function \\
			\( e_{jk} \) & Weight associated with GPU \( k \) of PM\(_j\) in the objective function \\
			\( m_i \) & Binary indicator of inter-machine migration of VM\(_i\) \\
			\( \omega_i \) & Binary indicator that a GI of VM\(_i\) changed GPU or slice \\
			\( \delta_i \) & Migration penalty of VM\(_i\), and zero for newly arrived VMs \\
			\( c_i \) & CPU requirement of VM\(_i\) \\
			\( C_j \) & CPU capacity of PM\(_j\) \\
			\( r_i \) & RAM requirement of VM\(_i\) \\
			\( R_j \) & RAM capacity of PM\(_j\) \\
			\( V_i \) & Set of GIs requested by VM\(_i\) \\
			\( p_v \) & Profile of GI \( v \) \\
			\( g_v \) & Size of GI \( v \) \\
			\( \mathcal{S}_{ivjk} \) & Permissible start slices of GI \( v \) of VM\(_i\) on GPU \( k \) of PM\(_j\) \\
			\( \varphi_j \) & Binary variable indicating if PM\(_j\) is powered-on \\
			\( \gamma_{jk} \) & Binary variable indicating if GPU \( k \) of PM\(_j\) hosts a GI \\
			\( H_{jk} \) & Compatibility class of the \( k^{th} \) GPU in PM\(_j\) \\
			\( h_v \) & Compatibility class of GI \( v \) \\
			\( o_{jku} \) & Indicates whether slice \( u \) of GPU \( k \) of PM\(_j\) is occupied \\
			\( \mathcal{P}_{jk} \) & Profiles supported by GPU \( k \) of PM\(_j\) \\
			\( w_p \) & Weight of a free start slice of profile \( p \) \\
			\( f_{jkpz} \) & Indicates if start slice \( z \) of GPU \( k \) is free for profile \( p \) \\
			\hline
		\end{tabular}
		\label{tab:notations}
	\end{table}

\begin{table}[ht]
	\centering
	\caption{Values for A100 40GB MIG profiles; other GPU families supply their own.}
	\begin{tabular}{llll}
		\hline
		Profile & $g_v$ & $s_p$ & $h_v$ \\ \hline
		1g.5gb  & 1     & 6     & 100 \\
		1g.10gb & 2     & 6     & 100 \\
		2g.10gb & 2     & 4     & 100 \\
		3g.20gb & 4     & 4     & 100 \\
		4g.20gb & 4     & 0     & 100 \\
		7g.40gb & 8     & 0     & 100 \\ \hline
	\end{tabular}
	\label{tab:model-params}
\end{table}

	Table \ref{tab:model-params} presents example parameters for various NVIDIA A100 MIG profiles. In this illustration, it is assumed that \( H_{jk} = 100 \) for every \( j \in M \) and \( k \in K_j \), when \( k \) corresponds to an A100 GPU. The collection \( U_{jk} \) then holds the 8 memory slices of the GPU. The profiles with identical memory requirements are given similar \( g_v \) values, although their compute slice needs may vary. These example settings of \( g_v \) and \( s_p \) demonstrate how the allocation can be managed.

	The weights \( a_i \) can be used to rank VMs based on predefined criteria. For example, assigning them equal values leaves the tendency to accept smaller VMs preferentially, since a GPU can host several small GIs in place of one large GI. Higher \( a_i \) values for larger GPU demands mitigate this tendency. In contrast, the weights \( b_j \) can modify the selection priority of physical machines. For example, higher values for machines with fewer GPUs improve consolidation.
	
	The parameter \( \delta_i \), set to zero for new VMs, omits their initial placement from the migration count. Conversely, a higher value inhibits the migration of VMs of interest. For instance, an elevated value assigned to a previously migrated VM reduces the likelihood of frequent VM preemptions. \( \delta_i \) sets the cost of a migration rather than the number permitted, as Eq.~\eqref{eq:1} penalizes migrations without bounding their total. A penalty rather than a bound lets the model migrate a VM whenever the resulting gain in acceptance or consolidation outweighs \( \delta_i \). Moreover, inter-machine migrations (\(m_i = 1\)) of VMs holding a GI also flip \(\omega_i = 1\), but intra-machine migrations do not result in machine switches. This model characteristic halves the penalty of intra-machine migrations relative to inter-machine migrations.

	\emph{Complexity.} The size of the BILP depends on the number of VMs and on the size of the infrastructure. The model contains \( |N||M| \) variables \( x_{ij} \). In addition, each GI has one variable \( y_{ivjkz} \) for every permissible start slice on every GPU of its compatibility class, which is at most \( |U_{jk}| \) per GPU. As a result, the number of variables, constraints, and nonzero coefficients is proportional to \( |N| \) multiplied by the number of PMs and GPUs, \( |M| + \sum_{j \in M} |K_j| \), and the model is built in time proportional to its size. Solving the model is much harder than building it. The problem includes bin packing as a special case, and bin packing is strongly NP-complete \cite{garey1979computers}. Consequently, the BILP is strongly NP-hard, and no exact algorithm can solve it in polynomial time unless P = NP. In practice, the solver finds a good placement quickly, because keeping all VMs in place and rejecting the new requests is always a feasible solution. Most of the time is spent on proving that a placement is optimal, and this proof is expensive when many PMs and GPUs are identical.

	\section{The Proposed Method}
	\label{s:methodology}
	
	In this section, we present OrigaMIG. As Section~\ref{s:formulation} explains, the BILP is strongly NP-hard, and proving a placement optimal becomes expensive as the data center grows. OrigaMIG therefore solves the BILP for a few PMs at a time and keeps the rest of the data center fixed. It has three components. An arrival rule places each request immediately, and triggers decide when the optimizer runs. The optimizer selects a neighborhood of PMs and solves the BILP on it. An executor turns the solution into a valid order of live migrations. OrigaMIG never removes a running VM.

	\subsection{Arrival Rule and Triggers}
	\label{s:dmig-triggers}

	OrigaMIG places each arriving request immediately, without a solver. It selects the first PM that can host the VM, and the default GI placement policy of Section~\ref{s:default-placement-policy} places each GI on its GPU. A VM placed by this arrival rule becomes a running VM from then on. A request that the arrival rule cannot place joins the pending set \( Q \). An admission call starts whenever \( Q \) is not empty, and a consolidation call starts when the threshold below is reached. We also allow a timer to start consolidation calls periodically, but we leave it off because the threshold already reacts to fragmentation. Only one call runs at a time, and admission calls take priority.

	For a compatibility class \( h \), the threshold uses the gap
	\begin{align}
		\notag \Gamma_h = & \sum_{j \in M} \sum_{k \in K_j,\, H_{jk} = h} \gamma_{jk} \\
		\label{eq:dmig-gap} & - \Big\lceil \frac{1}{|U_h|} \sum_{i \in N} \sum_{v \in V_i,\, h_v = h} g_v \Big\rceil,
	\end{align}
	where the sum over \( i \) runs over the placed VMs, and \( |U_h| \) is the number of memory slices of a GPU of class \( h \). The first term counts the active GPUs of class \( h \). The second term is a lower bound on the number of GPUs of class \( h \) needed by the placed GIs of this class. The gap \( \Gamma_h \) therefore bounds the number of GPUs that consolidation can switch off, and the gap grows with fragmentation. We keep both sums as counters, so \( \Gamma_h \) is updated in constant time. A consolidation call for class \( h \) starts when \( \Gamma_h \geq \theta \). The threshold \( \theta \) is at least one, because \( \Gamma_h = 0 \) proves that no GPU can be switched off. A call that changes nothing disables the threshold of class \( h \) until \( \Gamma_h \) increases or a VM departs from a PM of class \( h \). A PM of class \( h \) is a PM with at least one GPU of class \( h \). We also treat the PMs without GPUs, \( M_0 \), as one more class. For this class, the gap is
	\begin{align}
		\label{eq:dmig-gap0} \Gamma_0 = & \sum_{j \in M_0} \varphi_j - \Big\lceil \max\Big( \frac{\sum_{i \in N_0} c_i}{C_{\max}}, \frac{\sum_{i \in N_0} r_i}{R_{\max}} \Big) \Big\rceil,
	\end{align}
	where \( N_0 \) holds the placed VMs that run on \( M_0 \), and \( C_{\max} \) and \( R_{\max} \) are the largest CPU and RAM capacities of a PM of \( M_0 \). The second term is a lower bound on the number of PMs of \( M_0 \) needed by the VMs of \( N_0 \).

	\subsection{Neighborhood Selection}
	\label{s:dmig-neighborhood}

	Each call selects a neighborhood \( \hat{M} \subseteq M \) of at most \( M_{\max} \) PMs. The set \( \hat{N} \) holds the VMs that run on \( \hat{M} \) and the pending VMs of the call. The call skips a PM whose VMs would raise \( |\hat{N}| \) above \( N_{\max} \). The only exception is the first PM selected for the oldest pending VM, which ensures that every admission call reaches a decision for that VM. The limits \( M_{\max} \) and \( N_{\max} \) keep the restricted BILP small enough to be solved to optimality within seconds.

	OrigaMIG relies on four relations among the weights of Eq.~\eqref{eq:1}. First, \( a_i \) exceeds \( b_j \) plus \( e_{jk} \) for each GI of the VM, so the admission of one VM is worth switching on a PM and the GPUs that its GIs use. Second, \( b_j \) and \( e_{jk} \) exceed the migration penalties of a few VMs, so switching off a PM is worth a few migrations. Third, \( e_{jk} \) exceeds \( w_p \) times the \( \CF \) of an empty GPU of its class, which Section~\ref{s:default-placement-policy} defines. As a result, a GPU that becomes active always costs more than the free start slices it adds. Fourth, \( \delta_i \) of a placed VM is large compared with \( w_p \). A migration made only to raise \( \CF \) must then add more than \( \delta_i / w_p \) free start slices. OrigaMIG therefore does not select PMs by \( \CF \). Instead, each neighborhood aims at an admission or at switching off PMs.

	\begin{algorithm}[t]
	\SetKwProg{Proc}{procedure}{}{}
	\SetKwFunction{Admit}{AdmissionCall}
	\SetKwFunction{Consolidate}{ConsolidationCall}
	\caption{Neighborhood selection of OrigaMIG}\label{alg:dmig-call}
	\Proc{\Admit{\( Q \)}}{
		\( \hat{M} \leftarrow \emptyset \)\;
		\For{each \( \text{VM}_i \in Q \), oldest first}{
			\( C \leftarrow \) PMs whose free resources fit \( \text{VM}_i \)\;
			\eIf{\( C \neq \emptyset \)}{
				add \( C \) to \( \hat{M} \), fewest VMs first\;
			}{
				\( C \leftarrow \) PMs that would fit \( \text{VM}_i \) if empty\;
				\eIf{\( C \neq \emptyset \)}{
					add \( C \) to \( \hat{M} \), by free memory slices\;
				}{
					reject \( \text{VM}_i \)\;
				}
			}
		}
		fill \( \hat{M} \) with active PMs, by free memory slices, then CPU\;
		add a powered-off PM that fits the oldest VM of \( Q \), if any\;
		\( \hat{N} \leftarrow \) VMs on \( \hat{M} \), plus VMs of \( Q \) whose first PM is in \( \hat{M} \)\;
	}
	\Proc{\Consolidate{\( h \)}}{
		\( S \leftarrow \) active PMs of class \( h \), fewest VMs first\;
		keep \( j \in S \) only if \( b_j + \sum_{k \in K_j,\, \gamma_{jk} = 1} e_{jk} > w_p \, \CF_j + \sum_{i \in N_j} 2\delta_i \)\;
		\( R \leftarrow \) other active PMs of class \( h \), plus active PMs of \( M_0 \)\;
		keep at most \( \tfrac{1}{2} M_{\max} \) PMs in \( S \); fill \( \hat{M} \) with \( S \), then \( R \)\;
		\While{some PM of \( S \) cannot be emptied into \( R \)}{
			remove the last PM from \( S \)\;
		}
		\( \hat{N} \leftarrow \) VMs on \( \hat{M} \)\;
	}
	
	\end{algorithm}

\emph{Admission call.} The selection of an admission call is in Algorithm~\ref{alg:dmig-call}. The free resources of a PM fit a VM when the PM has enough free CPU and RAM and each GI of the VM fits one of its GPUs. A test takes the GIs one by one, starting from the largest, and a GI fits a GPU when the GPU has enough free memory slices and compute slices for the profile of the GI, and Table~\ref{tab:a100-gis} lists the profiles of the A100. The test ignores start slices, which a rearrangement can change. The powered-off PM is added because it can receive displaced VMs, and it can host the pending VM when the arrival rule failed only because it places the GIs one at a time. The call rejects a pending VM if the solution does not admit it, and the other pending VMs wait for the next call.

	\emph{Consolidation call.} The selection of a consolidation call is in Algorithm~\ref{alg:dmig-call}. The sources \( S \) are the PMs that the call tries to empty, and the receivers \( R \) take their VMs. In the algorithm, \( \CF_j \) is the sum of the \( \CF \) of the active GPUs of \( j \), and \( N_j \) the VMs running on \( j \). A VM that leaves a PM costs up to \( 2\delta_i \), because its move sets both \( m_i \) and \( \omega_i \). The PMs without GPUs join the receivers because a VM without GIs may move to them in every consolidation call.

	\subsection{Restricted BILP}
	\label{s:dmig-bilp}

	Each call acts as one interval of Section~\ref{s:formulation}, with \( \hat{M} \) and \( \hat{N} \) in place of \( M \) and \( N \). This fixes all other PMs and VMs, and it sets \( x_{ij} = 0 \) for every VM of \( \hat{N} \) and every PM outside \( \hat{M} \). The primed variables hold the placement at the start of the call.

	The solver starts from the current placement with all pending VMs rejected, which satisfies every constraint. It runs multithreaded, and it stops at optimality or when a time limit is reached. OrigaMIG then evaluates Eq.~\eqref{eq:1} over \( \hat{M} \) and \( \hat{N} \) for the returned placement and for the current placement. It recomputes \( \varphi_j \), \( \gamma_{jk} \), \( m_i \), \( \omega_i \), and \( f_{jkpz} \) from each placement, because a solution found before the proof of optimality may leave them at other values. It also subtracts \( \delta_i \) for each parking move of \( \text{VM}_i \) that the executor plans within a PM, and \( 2\delta_i \) for each parking move to another PM. OrigaMIG applies the plan only if its value is strictly larger than the value of the current placement.

	Fixing most of a solution and solving the rest exactly is large neighborhood search \cite{shaw1998using,pisinger2010large}. Each call is one step of this search on the BILP. As Section~\ref{s:formulation} explains, the size of the restricted BILP is proportional to \( |\hat{N}| \) multiplied by \( |\hat{M}| + \sum_{j \in \hat{M}} |K_j| \), which \( M_{\max} \) and \( N_{\max} \) bound, so it stays bounded as the data center grows.

	\subsection{Execution of Migrations}
	\label{s:dmig-executor}

	The solution is a target state, not an order of actions. As Section~\ref{s:introduction} explains, the destination GI must exist before a live migration. In addition, the source GI stays until the migration completes. The executor therefore plans the steps on a copy of the state before it acts. Each step creates the new GIs of one VM at their start slices through the NVIDIA management library. It then migrates or starts the VM and destroys its old GIs. A step is ready when its new slices are free. A start or a move between PMs also needs free CPU and RAM on the destination. If no step is ready, the remaining steps block each other in a cycle, which may consist of a single step. The executor then parks the VM whose removal makes the most steps ready. It moves this VM to free memory and compute slices of its class and to free CPU and RAM that no remaining step needs. A parked VM blocks no step again, so each VM is parked at most once. For example, in Fig.~\ref{fig:a100-frag}(c), parking the VM of the 3g.20gb GI on another GPU unblocks all steps. If no parking place exists, OrigaMIG discards the plan, and no step has run. Otherwise, OrigaMIG switches on the PMs that the plan needs before the first step, and it powers off the PMs left empty after the last step. While a call runs, the arrival rule places no request on \( \hat{M} \), and a departure only removes steps.

	\subsection{Properties and Cost}
	\label{s:dmig-properties}

	Every intermediate state satisfies Eq.~\eqref{eq:4}, Eq.~\eqref{eq:5}, and Eq.~\eqref{eq:10}, because each step uses only free resources and a migrating VM holds resources at both ends. Each constraint of the BILP involves one PM, one GPU, or one VM, together with the \( x_{ij} \) and \( y_{ivjkz} \) that link them. No VM of \( \hat{N} \) is placed outside \( \hat{M} \), and no other VM runs on \( \hat{M} \). Therefore, a feasible solution of the restricted BILP is feasible for the full BILP when combined with the fixed part of the data center, and the two objectives differ by a constant. This holds for any \( \hat{M} \), so the guarantees below do not depend on the selection rules.

	OrigaMIG offers three guarantees. First, a call never lowers the value of Eq.~\eqref{eq:1}, since a plan is applied only if it strictly improves the current placement after parking moves are counted. Second, if the solver proves optimality and the plan is applied, the placement is optimal for Eq.~\eqref{eq:1} among all placements that keep every running VM, admit only the pending VMs of \( \hat{N} \), and change only \( \hat{M} \). Third, without arrivals and departures, the calls stop changing the data center. The migration terms of Eq.~\eqref{eq:1} are never positive, so each applied plan strictly raises the other terms. These terms take only finitely many values. The time limit can invalidate only the second guarantee. However, OrigaMIG does not guarantee optimality for the whole data center or over time, since the selection rules are heuristics and the setting is non-clairvoyant. Finally, a call costs one pass over the PMs, solving within the time limit, and at most \( 2|\hat{N}| \) steps.
	
	\section{Evaluation}
	\label{s:evaluation}
	
	We evaluate OrigaMIG in simulation against online placement policies and against the optimum of the BILP. The evaluation answers four questions. First, does OrigaMIG admit more requests, keep fewer PMs and GPUs active, and use less energy than the baselines, across loads, workloads, and hardware (Sections~\ref{s:eval-comparison} and \ref{s:eval-robustness})? Second, what do its migrations cost (Section~\ref{s:eval-migration})? Third, what does each of its components contribute, and does OrigaMIG keep up as the data center grows (Section~\ref{s:eval-ablation})? Fourth, how far is it from the optimum, and how does the cost of finding the optimum grow (Sections~\ref{s:optgap} and \ref{s:bilp-scalability})?

	\subsection{Experimental Setup}
	\label{s:setup}
	
	We evaluate the placement policies in GPUCloudSim \cite{siavashi2019gpucloudsim}. The simulated PMs carry up to two MIG-enabled GPU models, the A100 40GB and the A30 24GB. Their profiles and values in the BILP are in Tables~\ref{tab:a100-gis}, \ref{tab:model-params}, and \ref{tab:a30-gis}. Each model forms its own compatibility class (\( H_{jk} = 100 \) for the A100, \( H_{jk} = 30 \) for the A30), so a GI runs only on the model of its profile.

	\begin{table}[t]
		\centering
		\scriptsize
		\caption{GPU instance profiles on A30 24GB and their values in the BILP}
		\begin{tabular}{|c|c|c|c|c|c|c|c|}
			\hline
			\textbf{Profile} & \textbf{Compute} & \textbf{Memory} & \textbf{Instances} & \textbf{Start Slices} & \( g_v \) & \( s_p \) & \( h_v \) \\ \hline
			1g.6gb  & 1/4 & 1/4 & 4 & 0, 1, 2, 3 & 1 & 3 & 30 \\ \hline
			2g.12gb & 2/4 & 2/4 & 2 & 0, 2       & 2 & 2 & 30 \\ \hline
			4g.24gb & 4/4 & 4/4 & 1 & 0          & 4 & 0 & 30 \\ \hline
		\end{tabular}
		\label{tab:a30-gis}
	\end{table}

	The BILP of Section~\ref{s:formulation} is modeled and solved with the CP-SAT engine of Google OR-Tools \cite{ortools}. In each OrigaMIG call, the solver runs on eight threads with a time limit of 10~s, and the neighborhood holds at most \( M_{\max} = 16 \) PMs and \( N_{\max} = 64 \) VMs. The consolidation threshold is \( \theta = 1 \), the smallest value that Section~\ref{s:dmig-triggers} allows. Every 10\,s, OrigaMIG checks whether to start a call.

	The weights are \( a_i = 100 \), \( b_j = 50 \), \( e_{jk} = 20 \), and \( w_p = 1 \) for every profile, and the migration penalty is \( \delta_i = 10 \) for a placed VM and 0 for a new one. They satisfy the four relations of Section~\ref{s:dmig-neighborhood}. First, a VM requests at most two GIs, so \( a_i \) exceeds \( b_j + 2e_{jk} = 90 \), the cost of switching on a PM and two GPUs. Second, \( b_j \) exceeds \( 4\delta_i = 40 \), the cost of two moves between PMs, and \( e_{jk} \) exceeds \( \delta_i = 10 \), the cost of one move within a PM. Third, \( e_{jk} \) exceeds the \( \CF \) of an empty A100, 18, and of an empty A30, 7. Fourth, a migration made only to raise \( \CF \) must gain more than \( \delta_i/w_p = 10 \) in \( \CF \), more than half the \( \CF \) of an empty A100, so OrigaMIG migrates only to admit a VM or to switch off a PM or a GPU.

	In Sections~\ref{s:eval-comparison}--\ref{s:eval-ablation}, unless stated otherwise, the data center has 128 PMs, 96 with two A100 GPUs and 32 with two A30 GPUs. Each PM has 80 cores and 512 GB of RAM. With 128 PMs, the data center is about the size of one partition, such as a row or a pod, the unit on which large-scale allocators optimize \cite{newell2021ras,narayanan2021pop}. We study larger data centers in Section~\ref{s:eval-ablation}. Each run has a warm-up of 48\,h, after which we measure for 3 days. We repeat each run with five seeds in Section~\ref{s:eval-comparison} and with three seeds in the other subsections. A seed gives every policy the same requests, so we compare two policies on the same seed before averaging over seeds. We report means with 95\% confidence intervals.

	We measure the following metrics. The \emph{acceptance} is the share of the requested memory slices that the data center admits, so that a large request counts more than a small one. The \emph{active PMs} and \emph{active GPUs} are the time averages of the numbers of PMs that host a VM and of GPUs that hold a GI. The \emph{packing density} is the share of the memory slices of the active GPUs that GIs occupy. The \emph{energy} is that consumed by the active PMs and their GPUs, and we also report it per admitted GPU-hour. The \emph{switchings} are the number of PMs switched on per hour, observed once per minute. For OrigaMIG and TC, we count the \emph{migrations} per hour, and, for OrigaMIG, the time that a request waits in \( Q \).

	We compute the energy as follows. We assume that a PM without VMs is switched off immediately and switched on when a VM needs it, and we ignore the time and energy of switching. An active PM draws the power that a SPECpower\_ssj2008 measurement of a server with the same CPUs reports for its load \cite{spec_dl380_8380}; the load is the share of its cores that the VMs use, and we interpolate linearly between the reported load levels. A GPU with GIs draws \( P_{\mathrm{idle}} + (P_{\max} - P_{\mathrm{idle}})\,u \), where \( u \) is the share of its compute slices that GIs occupy, as in GPUCloudSim \cite{siavashi2019gpucloudsim}. A GPU without GIs draws \( P_{\mathrm{idle}} \), since a GPU cannot power off while its PM is running \cite{nvidia_a30_brief}. Table~\ref{tab:power} lists the powers. The idle power of the A100 is measured on its SXM4 version \cite{cornelius2025telemetry}. We found no idle-power measurement for the A30, so we estimate its idle power by scaling its maximum power with the idle-to-maximum power ratio of the A100. In the energy per admitted GPU-hour, a GI counts as the fraction of a GPU equal to its share of the compute slices.

	\begin{table}[t]
		\centering
		\scriptsize
		\caption{Powers of the energy model}
		\begin{tabular}{|c|c|c|c|}
			\hline
			\textbf{Device} & \( P_{\mathrm{idle}} \) & \( P_{\max} \) & \textbf{Source} \\ \hline
			PM without its GPUs (2\( \times \) Xeon Platinum 8380) & 143 W & 647 W & \cite{spec_dl380_8380} \\ \hline
			A100 40GB PCIe & 50 W & 250 W & \cite{cornelius2025telemetry,nvidia_a100_brief} \\ \hline
			A30 24GB & 33 W & 165 W & \cite{nvidia_a30_brief} \\ \hline
		\end{tabular}
		\label{tab:power}
	\end{table}

	\subsection{Workload}
	\label{s:workload}

	To our knowledge, no public trace captures VM requests for MIG profiles. Existing GPU cluster traces, such as Alibaba's \cite{weng_beware_2023,siavashi_gpu_2024}, record jobs requesting whole GPUs or GPU fractions, which do not map directly to MIG profiles. We therefore generate synthetic workloads, combining transient jobs with the long-lived, heterogeneous tenants of self-hosted LLM serving \cite{pan2025costbenefit}. As a check, we map the Alibaba trace to MIG profiles in Section~\ref{s:eval-robustness}.

	Requests arrive according to a Poisson process. Of the requests, 70\% are jobs with a mean lifetime of 2\,h, and 30\% are tenants with a mean lifetime of 24\,h. Lifetimes are exponential. A VM requests one GI, or two GIs with probability 0.5. All its GIs are of one GPU model, and each profile of that model is equally likely. A VM also requests 4 cores and 16 GB of RAM per compute slice of its GIs. The lifetime of a VM starts when it is admitted. We set the arrival rate \( \lambda \) by the offered load \( \rho = \lambda\, \mathbb{E}[L]\, \mathbb{E}[f] / |K| \), where \( L \) is the lifetime of a VM, \( f \) the share of the memory slices of a GPU that the GIs of the VM occupy, and \( |K| \) the number of GPUs. The GPU model of a VM is drawn so that both models carry the same load \( \rho \). The workload of Section~\ref{s:optgap} has \( \rho = 0.74 \).
	
	\subsection{Baselines}
	\label{s:baselines}
	
	We compare OrigaMIG against the following placement policies. All of them are online and non-clairvoyant like OrigaMIG, and they place the GIs of a multi-GI VM one GI at a time.
	\begin{itemize}
		\item \emph{First-Fit} (FF) pairs first-fit host selection \cite{siavashi2019gpucloudsim} with the default GI placement policy of Section~\ref{s:default-placement-policy}: the VM is placed on the first host that can accommodate it, and each GI is placed at the start slice that maximizes \(\CF\), with ties broken by the lowest slice. FF is the default placement, and we report the gains of the other policies over it.
		\item \emph{Minimum Fragmentation Increment} (MFI) is the online policy of Zambianco et al.\ \cite{zambianco2025online}. It scores the fragmentation of a GPU by summing, for each profile whose memory slices fit in the free slices of the GPU, the number of memory slices of the profile times the number of its start slices at which the required slices are not all free. A GI goes to the GPU and start slice that increase this score the least.
		\item \emph{Turkkan Compaction} (TC) is the rule-based placement and compaction of Turkkan et al.\ \cite{turkkan2024optimal}. A GI goes to the GPU whose joint slice utilization \( (s_m + s_c)/(S_m + S_c) \) after the assignment is the largest, at a start slice taken from a fixed preference order for the A100. Here \( s_m \) and \( s_c \) are the used memory and compute slices of the GPU, and \( S_m \) and \( S_c \) are its memory and compute slices. Periodically, TC examines the active GPUs, starting from the least utilized, and vacates a GPU when all its workloads fit on the other active GPUs by the same rule. The moves use only resources that were free before the compaction, so they apply at once. If a GPU cannot be vacated, TC adds a free GPU to the compaction and keeps the moves only if they vacate at least two GPUs.
	\end{itemize}

	\emph{Adaptations.} MFI and TC were evaluated with single GIs on GPUs of one model, without CPU and RAM. We adapt both to VMs on PMs and otherwise keep their rules. In both, the GIs of a VM, largest first, go to one PM with enough CPU and RAM, so a VM moves as a whole. A GI goes only to GPUs of its class. Ties are broken by the lowest PM, GPU, and start slice, and a VM that fits nowhere is rejected. MFI computes the score of a GPU with the profiles and start slices of its model and chooses the PM with the smallest total increase. TC chooses the PM with the largest mean utilization of the GPUs that the VM uses. The original TC places a batch of new workloads and keeps those that do not fit pending. We place each VM when it arrives. We use the start slices of the default GI placement policy, which also covers the A30. TC compacts each class every 15\,min, a period the original leaves open. It takes the idle GPU from an active PM when one has an idle GPU, so that a compaction does not switch on a PM. Each VM moves at most once in a compaction, and a GPU that receives a VM is not vacated later in it. The executor of Section~\ref{s:dmig-executor} runs the moves of TC as live migrations, at the same cost as those of OrigaMIG, and no request is placed on the PMs of a compaction until its moves end.

	In addition, to establish an upper bound, we report \emph{Optimal} (OPT), which solves the BILP of Section~\ref{s:formulation} over the whole data center from a given state, which bounds the value any placement can reach from that state (Section~\ref{s:optgap}).
	Of the online policies compared, only OrigaMIG and TC migrate. The other works in Table~\ref{tab:related-work} are not used as baselines because they schedule jobs or inference services on homogeneous clusters or single nodes, whereas OrigaMIG places VMs across a heterogeneous, multi-host data center.

	\begin{figure}[t]
		\centering
		\begin{tikzpicture}
			\begin{groupplot}[
				group style={group size=1 by 2, vertical sep=2.05cm},
				width=\columnwidth, height=0.5\columnwidth,
				xmode=log, log basis x=2, ymode=log, ymin=0.2, ymax=5000,
				ytick={1,10,100,1000},
				ylabel={Time to optimality (s)},
				grid=major, grid style={black!10},
				tick label style={font=\scriptsize}, label style={font=\footnotesize}, title style={font=\footnotesize},
				legend style={font=\tiny, fill=white, fill opacity=0.85, text opacity=1, draw=black!30},
				every axis plot/.append style={mark size=1.6pt, line width=0.7pt},
				unbounded coords=discard,
			]
				\nextgroupplot[
					title={(a) Increasing PMs},
					xlabel={PMs \( |M| \) (top) and GPUs (bottom)},
					xtick={16,32,64,128,256},
					xticklabels={16\\32,32\\64,64\\128,128\\256,256\\512},
					xmin=12, xmax=340,
					xticklabel style={align=center},
					legend to name=bilplegend, legend columns=3, legend style={/tikz/every even column/.append style={column sep=4pt}},
				]
				\addplot[black, mark=*, error bars/.cd, y dir=both, y explicit] table[x=hosts, y=tproof, y error plus expr=\thisrow{tproofhi}-\thisrow{tproof}, y error minus expr=\thisrow{tproof}-\thisrow{tprooflo}] {\bilpMSweep};
				\addlegendentry{\( |N| = 64 \)}
				\addplot[black!50, dashed, mark=o, skip coords between index={0}{3}, error bars/.cd, y dir=both, y explicit] table[x=hosts, y=tproof, y error plus expr=\thisrow{tproofhi}-\thisrow{tproof}, y error minus expr=\thisrow{tproof}-\thisrow{tprooflo}] {\bilpGrow};
				\addlegendentry{\( |N| \) grows with \( |M| \)}
				\addplot[red!70!black, only marks, mark=x, mark size=2.4pt, point meta=explicit, nodes near coords={\pgfmathprintnumber[fixed, precision=1]{\pgfplotspointmeta}\%}, every node near coord/.append style={font=\tiny, anchor=south}] table[x=hosts, y=tcap, meta=gap] {\bilpMSweep};
				\addlegendentry{Limit reached, final gap}
				\addplot[red!70!black, only marks, mark=x, mark size=2.4pt, point meta=explicit, nodes near coords={\pgfmathprintnumber[fixed, precision=1]{\pgfplotspointmeta}\%}, every node near coord/.append style={font=\tiny, anchor=north west, inner sep=1pt}, forget plot] table[x=hosts, y=tcap, meta=gap] {\bilpGrow};
				\addplot[black!45, densely dashed, domain=12:340, samples=2, forget plot] {\bilpLimit};
				\node[font=\tiny, black!60, anchor=south west, inner sep=1pt] at (axis cs:12,\bilpLimit) {limit (\bilpLimitMin\,min)};
				\addplot[black!45, dotted, line width=0.9pt, domain=16:256, samples=2] {\bilpMProofSixteen*x/16};
				\addlegendentry{Model size}
				\addlegendimage{blue!70!black, mark=triangle*}
				\addlegendentry{\( |M| = 32 \)}
				\bilpMSteps

				\nextgroupplot[
					title={(b) Increasing VMs},
					xlabel={VMs \( |N| \) on \( |M| = 32 \) PMs (64 GPUs)},
					xtick={8,16,32,64,128}, xticklabels={8,16,32,64,128},
					xmin=6, xmax=170,
				]
				\addplot[blue!70!black, mark=triangle*, error bars/.cd, y dir=both, y explicit] table[x=nvms, y=tproof, y error plus expr=\thisrow{tproofhi}-\thisrow{tproof}, y error minus expr=\thisrow{tproof}-\thisrow{tprooflo}] {\bilpNSweep};
				
				\addplot[red!70!black, only marks, mark=x, mark size=2.4pt, point meta=explicit, nodes near coords={\pgfmathprintnumber[fixed, precision=1]{\pgfplotspointmeta}\%}, every node near coord/.append style={font=\tiny, anchor=south}, forget plot] table[x=nvms, y=tcap, meta=gap] {\bilpNSweep};
				\addplot[black!45, densely dashed, domain=6:170, samples=2, forget plot] {\bilpLimit};
				\node[font=\tiny, black!60, anchor=south west, inner sep=1pt] at (axis cs:6,\bilpLimit) {limit (\bilpLimitMin\,min)};
				\addplot[black!45, dotted, line width=0.9pt, domain=8:128, samples=2] {\bilpNProofEight*x/8};
				\bilpNSteps
			\end{groupplot}
			\node[anchor=south] at ([yshift=0.55cm]group c1r1.north) {\pgfplotslegendfromname{bilplegend}};
		\end{tikzpicture}
		\caption{Time to solve one interval to optimality as the numbers of PMs and VMs increase. Each point is the median of three seeds, and the error bars show the minimum and the maximum. The dotted line shows how the time would grow if it only doubled at each step, as the size of the model does. In (a), the dashed line keeps 3/4 of the memory slices occupied, so that the number of VMs grows with the number of PMs. A cross marks a point at which most solves reached the time limit, together with the median final gap.}
		\label{fig:bilp-scaling}
	\end{figure}
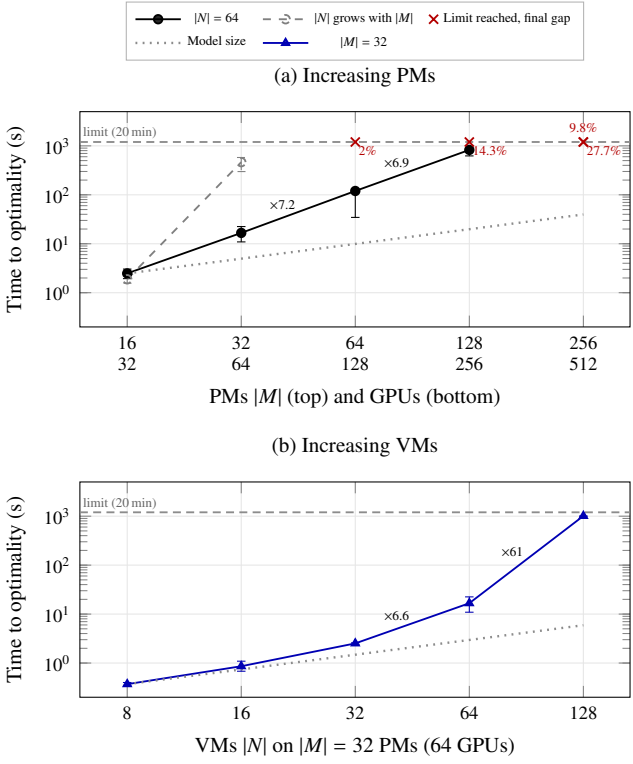

	\subsection{Comparison with the Baselines}
	\label{s:eval-comparison}

\begin{figure*}[t]
		\centering
		\begin{tikzpicture}
			\begin{groupplot}[
				group style={group size=4 by 1, horizontal sep=1.35cm},
				width=0.27\textwidth, height=0.22\textwidth,
				xmin=0.45, xmax=1.25, xtick={0.5,0.65,0.75,0.85,1,1.2}, xticklabels={0.5,,0.75,,1.0,1.2},
				xlabel={Offered load},
				grid=major, grid style={black!10},
				tick label style={font=\scriptsize}, label style={font=\footnotesize}, title style={font=\footnotesize},
				every axis plot/.append style={line width=0.7pt, mark size=1.6pt},
				error bars/y dir=both, error bars/y explicit,
				legend style={font=\scriptsize, draw=black!30, legend columns=4, /tikz/every even column/.append style={column sep=6pt}},
			]
				\nextgroupplot[title={(a) Acceptance}, ylabel={Admitted slices (\%)}, legend to name=evalSoneLegend]
				\addplot[black!60, dashed, mark=o] table[x=load, y=acc, y error=acc_ci] {\evalSoneFF}; \addlegendentry{FF}
				\addplot[orange!85!black, mark=triangle] table[x=load, y=acc, y error=acc_ci] {\evalSoneMFI}; \addlegendentry{MFI}
				\addplot[green!45!black, mark=diamond] table[x=load, y=acc, y error=acc_ci] {\evalSoneTC}; \addlegendentry{TC}
				\addplot[blue!70!black, mark=*] table[x=load, y=acc, y error=acc_ci] {\evalSoneDMIG}; \addlegendentry{OrigaMIG}
				\nextgroupplot[title={(b) Active PMs}, ylabel={Share of the PMs (\%)}]
				\addplot[black!60, dashed, mark=o] table[x=load, y=pm, y error=pm_ci] {\evalSoneFF};
				\addplot[orange!85!black, mark=triangle] table[x=load, y=pm, y error=pm_ci] {\evalSoneMFI};
				\addplot[green!45!black, mark=diamond] table[x=load, y=pm, y error=pm_ci] {\evalSoneTC};
				\addplot[blue!70!black, mark=*] table[x=load, y=pm, y error=pm_ci] {\evalSoneDMIG};
				\nextgroupplot[title={(c) Packing density}, ylabel={Occupied slices (\%)}]
				\addplot[black!60, dashed, mark=o] table[x=load, y=pack, y error=pack_ci] {\evalSoneFF};
				\addplot[orange!85!black, mark=triangle] table[x=load, y=pack, y error=pack_ci] {\evalSoneMFI};
				\addplot[green!45!black, mark=diamond] table[x=load, y=pack, y error=pack_ci] {\evalSoneTC};
				\addplot[blue!70!black, mark=*] table[x=load, y=pack, y error=pack_ci] {\evalSoneDMIG};
				\nextgroupplot[title={(d) Energy per GPU-hour}, ylabel={kWh per admitted GPU-hour}, y tick label style={/pgf/number format/.cd, fixed, precision=3}]
				\addplot[black!60, dashed, mark=o] table[x=load, y=epw, y error=epw_ci] {\evalSoneFF};
				\addplot[orange!85!black, mark=triangle] table[x=load, y=epw, y error=epw_ci] {\evalSoneMFI};
				\addplot[green!45!black, mark=diamond] table[x=load, y=epw, y error=epw_ci] {\evalSoneTC};
				\addplot[blue!70!black, mark=*] table[x=load, y=epw, y error=epw_ci] {\evalSoneDMIG};
			\end{groupplot}
			\node[anchor=south] at ([yshift=0.75cm]$(group c2r1.north)!0.5!(group c3r1.north)$) {\pgfplotslegendfromname{evalSoneLegend}};
		\end{tikzpicture}
		\caption{FF, MFI, TC, and OrigaMIG on 128 PMs as the offered load grows, over \evalSoneSeeds{} seeds. The error bars show the 95\% confidence intervals.}
		\label{fig:eval-comparison}
	\end{figure*}
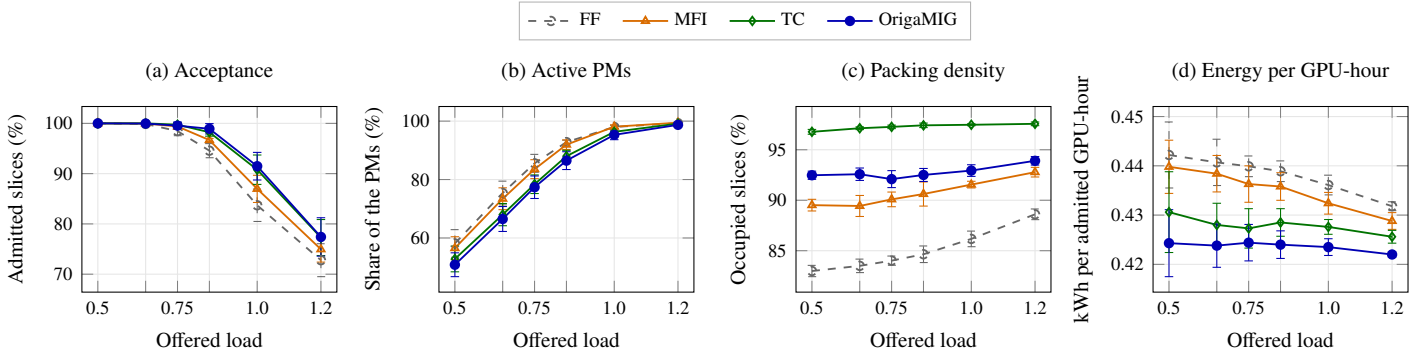

	We compare OrigaMIG with FF, MFI, and TC as the offered load grows from 0.5 to 1.2 (Fig.~\ref{fig:eval-comparison} and Table~\ref{tab:eval-comparison}). Each value is a mean over \evalSoneSeeds{} seeds, and Fig.~\ref{fig:eval-comparison} shows the 95\% confidence intervals.

	At low load, the data center admits nearly every request, so the policies differ in the PMs they keep active. At a load of 0.5, OrigaMIG keeps \evalSoneGainDMIGDpmZeropFive\% fewer PMs active than FF, TC \evalSoneGainTCDpmZeropFive\%, and MFI \evalSoneGainMFIDpmZeropFive\%. TC packs the active GPUs tighter, \evalSoneTCPackZeropFive\% against \evalSoneDMIGPackZeropFive\%, and keeps fewer GPUs active, because it vacates GPUs, whereas the weights of OrigaMIG value an empty PM more than an empty GPU. As the load grows, fewer PMs can be emptied, and the reduction of OrigaMIG falls to \evalSoneGainDMIGDpmZeropEightFive\% at a load of 0.85 and \evalSoneGainDMIGDpmOne\% at 1.0. Paired by seed, OrigaMIG keeps fewer PMs active than TC in \evalSoneVsTCWins{} of \evalSoneVsTCRuns{} runs, by \evalSoneVsTCDpmMin\% to \evalSoneVsTCDpmMax\% depending on the load.

	At high load, the requests compete for GPUs. At a load of 1.0, OrigaMIG admits \evalSoneDMIGAccOne\% of the requested memory slices, against \evalSoneFFAccOne\% for FF, \evalSoneMFIAccOne\% for MFI, and \evalSoneTCAccOne\% for TC. An admission call rearranges GIs to fit a request that the arrival rule could not place, and the admitted requests wait \evalSoneDMIGWaitOne\,s on average.

	As OrigaMIG and TC admit more work at high load, they also use more energy there than FF (Table~\ref{tab:eval-comparison}). Per admitted GPU-hour, OrigaMIG uses the least energy at every load, \evalSoneGainDMIGDepwZeropFive\% less than FF at a load of 0.5 and \evalSoneGainDMIGDepwOne\% less at 1.0. It migrates \evalSoneDMIGMigZeropFive{} VMs per hour at a load of 0.5 and \evalSoneDMIGMigOne{} at 1.0, against \evalSoneTCMigZeropFive{} and \evalSoneTCMigOne{} for TC, that is, \evalSoneVsTCDmigMin\% to \evalSoneVsTCDmigMax\% fewer across the loads (Section~\ref{s:eval-migration}). In exchange, it switches PMs on more often than the other policies.

	\begin{table*}[t]
		\centering
		\scriptsize
		\caption{Results at two offered loads, over \evalSoneSeeds{} seeds}
		\begin{tabular}{|c|c|c|c|c|c|c|c|c|c|}
			\hline
			\textbf{Load} & \textbf{Policy} & \textbf{Accept. (\%)} & \textbf{Active PMs} & \textbf{Active GPUs (\%)} & \textbf{Packing (\%)} & \textbf{Energy (MWh)} & \textbf{Switch./h} & \textbf{Migr./h} & \textbf{Wait (s)} \\ \hline
			\evalSoneRows
		\end{tabular}
		\label{tab:eval-comparison}
	\end{table*}

	\subsection{Robustness}
	\label{s:eval-robustness}

	We vary one factor of the workload or the hardware at a time, at a load of 0.85 (Table~\ref{tab:eval-variants}); the other factors keep the values of Sections~\ref{s:setup} and \ref{s:workload}. For the trace variant, we map the Alibaba GPU trace of 2023 \cite{weng_beware_2023} to A100 profiles. A pod that requests a share \( f \) of a GPU gets the smallest profile that has at least \( \lceil 7f \rceil \) compute slices and \( \lceil 8f \rceil \) memory slices. A pod that requests one or two GPUs gets one or two 7g.40gb GIs. We drop the pods without a GPU and those that request more than two. The trace ends with a list of the pods still alive, so we keep only the pods created in its last 34 days. We compress the timeline so that the offered load is 0.85, and seed \( s \) starts the trace at \( (s-1)/5 \) of its length. Of the resulting GIs, 77\% are 7g.40gb.

\begin{figure*}[t]
		\centering
		\begin{tikzpicture}
			\begin{groupplot}[
				group style={group size=2 by 1, horizontal sep=0.25cm, yticklabels at=edge left},
				width=0.62\columnwidth, height=7.2cm,
				xbar=0pt, /pgf/bar width=2.4pt,
				ymin=-0.6, ymax=11.6, y dir=reverse, ytick={0,...,11},
				yticklabels={Default, Smaller profiles, Larger profiles, One GI per VM, Two GIs per VM, Lognormal lifetimes, 10\% tenants, 60\% tenants, A100 PMs only, A30 PMs only, Fewer cores and RAM, Alibaba trace},
				xmajorgrids, grid style={black!10},
				tick label style={font=\scriptsize}, label style={font=\footnotesize}, title style={font=\footnotesize},
				error bars/x dir=both, error bars/x explicit, error bars/error bar style={black!50, line width=0.4pt}, error bars/error mark options={rotate=90, mark size=1pt},
				legend style={font=\scriptsize, draw=black!30, legend columns=3, /tikz/every even column/.append style={column sep=6pt}},
				legend image code/.code={\fill[#1, draw=none] (0cm,-0.06cm) rectangle (0.3cm,0.06cm);},
			]
				\nextgroupplot[title={(a) Acceptance}, xlabel={Gain (pp)}, legend to name=evalStwoLegend]
				\addplot[draw=none, fill=orange!85!black] table[y=v, x=dacc, x error=dacc_ci] {\evalStwoGainMFI}; \addlegendentry{MFI}
				\addplot[draw=none, fill=green!45!black] table[y=v, x=dacc, x error=dacc_ci] {\evalStwoGainTC}; \addlegendentry{TC}
				\addplot[draw=none, fill=blue!70!black] table[y=v, x=dacc, x error=dacc_ci] {\evalStwoGainDMIG}; \addlegendentry{OrigaMIG}
				\nextgroupplot[title={(b) Active PMs}, xlabel={Reduction (\%)}]
				\addplot[draw=none, fill=orange!85!black] table[y=v, x=dpm, x error=dpm_ci] {\evalStwoGainMFI};
				\addplot[draw=none, fill=green!45!black] table[y=v, x=dpm, x error=dpm_ci] {\evalStwoGainTC};
				\addplot[draw=none, fill=blue!70!black] table[y=v, x=dpm, x error=dpm_ci] {\evalStwoGainDMIG};
			\end{groupplot}
			\node[anchor=south] at ([yshift=0.55cm]$(group c1r1.north)!0.5!(group c2r1.north)$) {\pgfplotslegendfromname{evalStwoLegend}};
		\end{tikzpicture}
		\caption{Gains of MFI, TC, and OrigaMIG over FF under each variant of Table~\ref{tab:eval-variants}, at a load of 0.85, over three seeds. The error bars show the 95\% confidence intervals.}
		\label{fig:eval-robustness}
	\end{figure*}
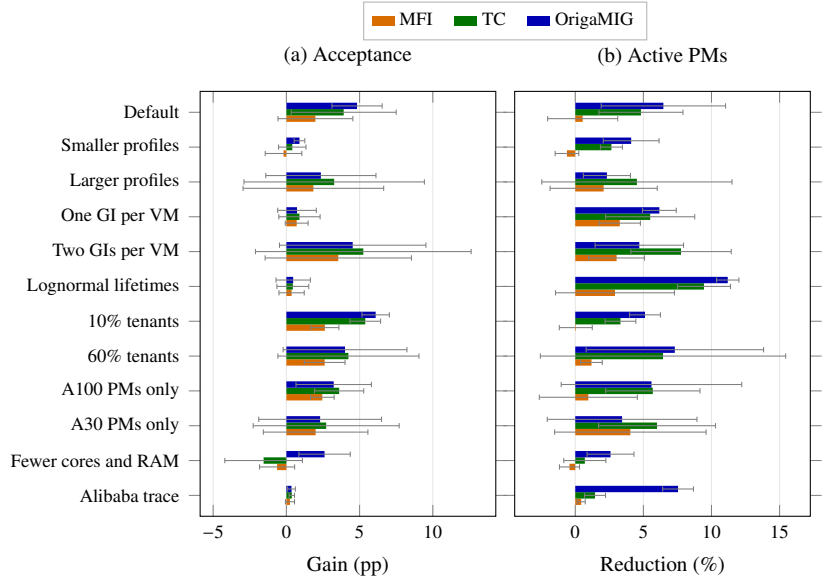

	Fig.~\ref{fig:eval-robustness} shows the gains of MFI, TC, and OrigaMIG over FF. The gains in acceptance follow the rejections of FF. With smaller profiles, one GI per VM, lognormal lifetimes, or the trace, FF admits more than 98\%, and no policy gains a full percentage point. With two GIs per VM and with 10\% tenants, FF admits \evalStwoFFAccTwoGi\% and \evalStwoFFAccTenantLow\%, and TC admits \evalStwoTCDaccTwoGi{} and \evalStwoTCDaccTenantLow{} percentage points more. TC keeps fewer PMs active than FF under every variant, and MFI gains less than TC under every variant but one.

	OrigaMIG keeps fewer PMs active than TC in seven of the eleven variants, about as many in one, and more in three. Its largest lead comes when the CPU and the RAM of the PMs run out before their GPUs. FF then admits only \evalStwoFFAccCpu\%, and neither MFI nor TC admits more, as both rank PMs by the state of their GPUs. The BILP of OrigaMIG also constrains the cores and the RAM (Eqs.~\eqref{eq:4} and \eqref{eq:5}), so OrigaMIG admits \evalStwoDMIGDaccCpu{} percentage points more than FF and keeps \evalStwoDMIGDpmCpu\% fewer PMs active, against \evalStwoTCDpmCpu\% for TC, with \evalStwoDMIGMigCpu{} migrations per hour against \evalStwoTCMigCpu. With smaller profiles, with one GI per VM, and with 10\% or 60\% tenants, OrigaMIG also keeps fewer PMs active than TC and migrates fewer VMs. With smaller profiles, for instance, it keeps \evalStwoDMIGDpmMixSmall\% fewer PMs active than FF, against \evalStwoTCDpmMixSmall\% for TC, with \evalStwoDMIGMigMixSmall{} migrations per hour against \evalStwoTCMigMixSmall. On the trace, most GIs take a whole GPU, so MFI and TC find little fragmentation to remove, and TC keeps \evalStwoTCDpmTrace\% fewer PMs active than FF. OrigaMIG keeps \evalStwoDMIGDpmTrace\% fewer with about as many active GPUs, at the cost of \evalStwoDMIGMigTrace{} migrations and \evalStwoDMIGGbhTrace\,GB per hour, against \evalStwoTCMigTrace{} and \evalStwoTCGbhTrace\,GB for TC. With A100 PMs only, the two keep about as many PMs active, \evalStwoDMIGDpmAhundred\% and \evalStwoTCDpmAhundred\% fewer than FF, and OrigaMIG migrates \evalStwoDMIGMigAhundred{} VMs per hour against \evalStwoTCMigAhundred. TC keeps fewer PMs active than OrigaMIG with larger profiles, with two GIs per VM, and with A30 PMs only, where the VMs are large relative to their GPUs. Its reductions over FF are \evalStwoTCDpmMixLarge\%, \evalStwoTCDpmTwoGi\%, and \evalStwoTCDpmAthirty\%, against \evalStwoDMIGDpmMixLarge\%, \evalStwoDMIGDpmTwoGi\%, and \evalStwoDMIGDpmAthirty\% for OrigaMIG (Section~\ref{s:discussion}).

	With lognormal lifetimes and the same means, many tenants outlive the simulated days, so the demand stays below the offered load. Over the measured days, the GIs occupy the memory slices of \evalStwoDemandLognormal{} GPUs, about as many as at a load of 0.65 in Setup 1 (\evalSoneDemandZeropSixFive). The reductions under this variant, \evalStwoDMIGDpmLognormal\% for OrigaMIG and \evalStwoTCDpmLognormal\% for TC, are close to those of Setup 1 at that load, \evalSoneGainDMIGDpmZeropSixFive\% and \evalSoneGainTCDpmZeropSixFive\%, so the shape of the lifetimes changes them little.

	\begin{table}[t]
		\centering
		\scriptsize
		\caption{Variants of the workload and the hardware}
		\begin{tabular}{|c|c|c|}
			\hline
			\textbf{Factor} & \textbf{Default} & \textbf{Variants} \\ \hline
			Profile mix & uniform & larger profiles half or twice as likely \\ \hline
			VMs with two GIs & 50\% & 0\%, 100\% \\ \hline
			Lifetimes & exponential & lognormal, \( \sigma = 1.5 \), same means \\ \hline
			Tenants & 30\% & 10\%, 60\% \\ \hline
			PMs & 96 A100, 32 A30 & 128 A100; 128 A30 \\ \hline
			Cores and RAM of a PM & 80, 512 GB & 32, 256 GB \\ \hline
			Requests & synthetic & Alibaba trace on 128 A100 PMs \\ \hline
		\end{tabular}
		\label{tab:eval-variants}
	\end{table}

	\subsection{Cost of Migrations}
	\label{s:eval-migration}

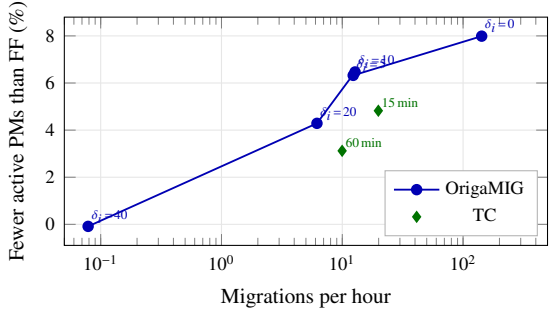
\begin{figure}[t]
		\centering
		\begin{tikzpicture}
			\begin{axis}[
				width=0.9\columnwidth, height=4.6cm,
				xmode=log, log basis x=10, xmin=0.05, xmax=500,
				xlabel={Migrations per hour}, ylabel={Fewer active PMs than FF (\%)},
				grid=major, grid style={black!10},
				tick label style={font=\scriptsize}, label style={font=\footnotesize},
				legend style={font=\scriptsize, draw=black!30, at={(0.98,0.04)}, anchor=south east},
				every axis plot/.append style={line width=0.7pt, mark size=1.8pt},
				every node near coord/.append style={font=\tiny, anchor=south west, inner sep=1pt},
			]
				\addplot[blue!70!black, mark=*, point meta=explicit, nodes near coords={\(\delta_i\!=\!\pgfmathprintnumber{\pgfplotspointmeta}\)}] table[x=mig, y=dpm, meta=delta] {\evalSthreeDMIG}; \addlegendentry{OrigaMIG}
				\addplot[green!45!black, mark=diamond*, only marks, point meta=explicit, nodes near coords={\pgfmathprintnumber{\pgfplotspointmeta}\,min}] table[x=mig, y=dpm, meta expr=\thisrow{period}/60] {\evalSthreeTC}; \addlegendentry{TC}
			\end{axis}
		\end{tikzpicture}
		\caption{Reduction in active PMs over FF against the migration rate, at a load of 0.85. The labels give the migration penalty \( \delta_i \) of OrigaMIG and the period of TC.}
		\label{fig:eval-migration}
	\end{figure}

	At a load of 0.85, we vary the migration penalty \( \delta_i \) of OrigaMIG from 0 to 40, and we run TC with a compaction every 15\,min or every hour (Fig.~\ref{fig:eval-migration} and Table~\ref{tab:eval-migration}). For \( \delta_i \geq 12.5 \), switching off a PM no longer outweighs the cost of two migrations between PMs, so the second relation of Section~\ref{s:dmig-neighborhood} no longer holds. A migration between PMs copies the RAM of the VM and the memory of its GIs at half the 25\,Gbit/s speed of a PM.

	The penalty sets the trade-off between migrations and active PMs. Without it, \( \delta_i = 0 \), OrigaMIG migrates \evalSthreeDeltaZeroMig{} VMs per hour, or \evalSthreeDeltaZeroMigvm{} per admitted VM, and copies \evalSthreeDeltaZeroGbh\,GB per hour, for \evalSthreeDeltaZeroDpm\% fewer PMs active than FF. With \( \delta_i = 5 \) and 10, it migrates \evalSthreeDeltaFiveMig{} and \evalSthreeDeltaTenMig{} VMs per hour, about a tenth as many, and keeps \evalSthreeDeltaFiveDpm\% and \evalSthreeDeltaTenDpm\% fewer PMs active. With \( \delta_i = 10 \), it copies \evalSthreeDeltaTenGbh\,GB per hour, or \evalSthreeDeltaTenMigvm{} migrations per admitted VM, and a migration between PMs takes \evalSthreeDeltaTenCopy\,s on average. With \( \delta_i = 20 \), OrigaMIG migrates \evalSthreeDeltaTwentyMig{} VMs per hour for \evalSthreeDeltaTwentyDpm\% fewer PMs, and with \( \delta_i = 40 \), it almost never migrates and keeps as many PMs active as FF. TC migrates \evalSthreeTCQuarterMig{} VMs per hour with a compaction every 15\,min, for \evalSthreeTCQuarterDpm\% fewer PMs, and \evalSthreeTCHourMig{} with a compaction every hour, for \evalSthreeTCHourDpm\%. For each period of TC, a value of \( \delta_i \) thus makes OrigaMIG migrate fewer VMs and keep fewer PMs active than TC (Fig.~\ref{fig:eval-migration}).

	The call times of Table~\ref{tab:eval-migration} depend on \( \delta_i \) through the neighborhoods. A consolidation call takes a PM as a source only if emptying it outweighs the penalties of its VMs (Algorithm~\ref{alg:dmig-call}). With \( \delta_i \leq 5 \), lightly loaded sources fill half of the neighborhood, and a call holds about 11 VMs, against 44 to 58 with \( \delta_i \geq 10 \).

	\begin{table*}[t]
		\centering
		\scriptsize
		\caption{Migrations and gains at a load of 0.85. The call time is the median runtime of a call}
		\begin{tabular}{|c|c|c|c|c|c|c|c|c|}
			\hline
			\textbf{Policy} & \textbf{Seeds} & \textbf{Migr./h} & \textbf{Migr. per VM} & \textbf{GB/h} & \textbf{Accept. (\%)} & \textbf{Active PMs} & \textbf{Energy (MWh)} & \textbf{Call (s)} \\ \hline
			\evalSthreeRows
		\end{tabular}
		\label{tab:eval-migration}
	\end{table*}

	\subsection{Ablation and Scale}
	\label{s:eval-ablation}

	We disable one component of OrigaMIG at a time and vary its parameters, at a load of 0.85 (Table~\ref{tab:eval-ablation}). Without admission calls, a request that the arrival rule cannot place is rejected. Without consolidation calls, OrigaMIG only admits pending VMs. The timer variant starts a consolidation call for each class every 15\,min instead of when the threshold is reached. The random variant selects the PMs of a call at random, within \( M_{\max} \) and \( N_{\max} \), instead of using Algorithm~\ref{alg:dmig-call}. Without any call, OrigaMIG reduces to FF. We vary \( M_{\max} \) from 4 to 32 with \( N_{\max} = 4M_{\max} \), the time limit from 1 to 30\,s, and \( \theta \) from 1 to 4.

	Both kinds of calls are needed. Without consolidation calls, OrigaMIG keeps \evalSfourNoConsPmabs{} PMs active, as many as FF with \evalSfourFFPmabs{}, and admits \evalSfourNoConsAcc\% of the requested slices instead of \evalSfourFullAcc\%. Without admission calls, it keeps as few PMs active but admits only \evalSfourNoAdmAcc\%. With the timer, a class may wait up to 15\,min for its next call, so OrigaMIG migrates half as many VMs but keeps \evalSfourTimerPmabs{} PMs active. Random neighborhoods do as well as Algorithm~\ref{alg:dmig-call}, with \evalSfourRandomPmabs{} active PMs against \evalSfourFullPmabs{}, so at this load, the gains come from the restricted BILP rather than from the choice of its PMs. The neighborhood must be large enough to empty a PM. With \( M_{\max} = 4 \), OrigaMIG keeps \evalSfourMmaxFourPmabs{} PMs active, as many as FF, and with \( M_{\max} = 8 \), \evalSfourMmaxEightPmabs. With \( M_{\max} = 32 \), the median call takes \evalSfourMmaxThirtyTwoCallmed\,s, so most calls stop at the time limit, and OrigaMIG keeps \evalSfourMmaxThirtyTwoPmabs{} PMs active, more than with \( M_{\max} = 16 \). The time limit matters up to 10\,s. A limit of 1\,s keeps \evalSfourLimitOnePmabs{} PMs active, 5\,s keeps \evalSfourLimitFivePmabs, and 30\,s keeps \evalSfourLimitThirtyPmabs, against \evalSfourFullPmabs{} with 10\,s. The threshold matters little. With \( \theta = 2 \) and 4, OrigaMIG keeps \evalSfourThetaTwoPmabs{} and \evalSfourThetaFourPmabs{} PMs active, and \( \theta = 4 \) lowers the calls from \evalSfourFullCalls{} to \evalSfourThetaFourCalls{} per hour.

\begin{figure}[t]
		\centering
		\begin{tikzpicture}
			\begin{groupplot}[
				group style={group size=2 by 1, horizontal sep=1.3cm},
				width=0.52\columnwidth, height=4.2cm,
				xmode=log, log basis x=2, xmin=100, xmax=5200, xtick={128,512,1024,2048,4096}, xticklabels={128,512,1k,2k,4k},
				xlabel={PMs},
				grid=major, grid style={black!10},
				tick label style={font=\scriptsize}, label style={font=\footnotesize}, title style={font=\footnotesize},
				every axis plot/.append style={line width=0.7pt, mark size=1.6pt},
				legend style={font=\scriptsize, draw=black!30, at={(0.97,0.03)}, anchor=south east},
			]
				\nextgroupplot[title={(a) Calls}, ylabel={Calls per hour}, ymin=0, ymax=400]
				\addplot[blue!70!black, mark=*] table[x=pms, y=calls] {\evalSfourScale};
				\addplot[black!45, densely dashed, domain=100:5200, samples=2] {360};
				\node[font=\tiny, black!60, anchor=south east, inner sep=1pt] at (axis cs:5000,360) {at most 360};
				\nextgroupplot[title={(b) Runtime of a call}, ylabel={Seconds}, ymin=0, ymax=12]
				\addplot[blue!70!black, mark=*] table[x=pms, y=callmed] {\evalSfourScale}; \addlegendentry{median}
				\addplot[blue!70!black, dashed, mark=o] table[x=pms, y=callp95] {\evalSfourScale}; \addlegendentry{95th pct.}
			\end{groupplot}
		\end{tikzpicture}
		\caption{OrigaMIG as the data center grows, at a load of 0.85, over one day and one seed.}
		\label{fig:eval-scale}
	\end{figure}
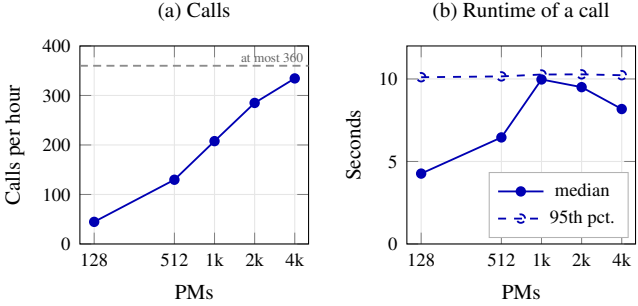

	We then grow the data center from 128 to 4096 PMs at the same load per PM, with a quarter of the PMs carrying A30 GPUs, and simulate one day with a single seed (Fig.~\ref{fig:eval-scale}). OrigaMIG starts at most one call per 10\,s step, that is, 360 calls per hour, while the calls it needs grow with the data center. The simulations shared one machine, so a call may take longer than on eight dedicated cores. From 128 to 1024 PMs, the calls grow from \evalSfourScaleCallsOneTwoEight{} to \evalSfourScaleCallsOneZeroTwoFour{} per hour, and the median call from \evalSfourScaleCallmedOneTwoEight\,s to \evalSfourScaleCallmedOneZeroTwoFour\,s, close to the time limit. From 1024 PMs on, FF admits every request, so OrigaMIG can gain only through fewer active PMs. Its reduction over FF falls from \evalSfourScaleDpmOneTwoEight\% with 128 PMs and \evalSfourScaleDpmFiveOneTwo\% with 512 PMs to \evalSfourScaleDpmOneZeroTwoFour\% with 1024 PMs, and it stays below 0.5\% with 2048 and 4096 PMs, where OrigaMIG starts \evalSfourScaleCallsTwoZeroFourEight{} and \evalSfourScaleCallsFourZeroNineSix{} calls per hour. One instance of OrigaMIG thus serves up to about 1024 PMs in our workload. A larger data center can run one instance per partition of its PMs, as large-scale allocators partition their optimization problems \cite{newell2021ras,narayanan2021pop}. Each constraint of the BILP involves one PM, one GPU, or one VM (Section~\ref{s:dmig-properties}), so calls on disjoint neighborhoods can also run in parallel.

	\begin{table*}[t]
		\centering
		\scriptsize
		\caption{Ablation and sensitivity of OrigaMIG at a load of 0.85}
		\begin{tabular}{|c|c|c|c|c|c|c|}
			\hline
			\textbf{Variant} & \textbf{Seeds} & \textbf{Accept. (\%)} & \textbf{Active PMs} & \textbf{Energy (MWh)} & \textbf{Migr./h} & \textbf{Calls/h} \\ \hline
			\evalSfourRows
		\end{tabular}
		\label{tab:eval-ablation}
	\end{table*}

	\subsection{Optimality Gap and Runtime}
	\label{s:optgap}

	OrigaMIG solves each call on a neighborhood of at most \( M_{\max}\nobreak\!=\!\nobreak16 \) PMs instead of the whole data center, trading solution quality for runtime. We quantify this trade-off against OPT, the optimal solution of the BILP over the whole data center, in terms of the optimality gap and the runtime of OrigaMIG and of OPT.

	\emph{Method.} We simulate OrigaMIG from an empty data center and take snapshots every 4\,h from 48\,h to 96\,h. At each snapshot, we solve the BILP over the whole data center from the current state, keeping every running VM placed, and compare its solution with the placement of OrigaMIG in Eq.~\eqref{eq:1}. OPT is the best placement from that state, so it upper-bounds the value OrigaMIG can reach in that state, but not over the whole run. The workload follows Section~\ref{s:workload}, with Poisson arrivals at 2.1 per hour per 8 PMs, 70\% transient jobs (mean 2\,h) and 30\% tenants (mean 24\,h), and one or two A100 GIs per VM. As OPT solves are costly, we run multiple seeds where feasible and report the worst-case gap. The solver proves the optimality of every OPT solve. The calls of OrigaMIG use the settings of Section~\ref{s:setup}.

	\begin{table}[t]
		\centering
		\scriptsize
		\caption{Optimality gap and runtime of OrigaMIG against OPT, with the settings of Section~\ref{s:setup} (\( M_{\max} = 16 \)). The optimality gap is given as the additional active PMs of OrigaMIG relative to OPT and as the relative gap in Eq.~\eqref{eq:1}, \( (\mathrm{OPT} - \mathrm{OrigaMIG}) / \mathrm{OPT} \). The runtime is the median wall-clock time of one OrigaMIG call and of one OPT solve.}
		\begin{tabular}{|c|c|c|c|c|c|}
			\hline
			& & \multicolumn{2}{c|}{\textbf{Optimality gap}} & \multicolumn{2}{c|}{\textbf{Runtime}} \\ \cline{3-6}
			\textbf{PMs} & \textbf{VMs} & \textbf{Active PMs} & \textbf{Eq.~\eqref{eq:1}} & \textbf{OrigaMIG call} & \textbf{OPT} \\ \hline
			8  & 19 & 3.9\% & 0.54\% & 0.3\,s & 0.3\,s \\ \hline
			16 & 30 & 4.1\% & 0.43\% & 0.9\,s & 1.7\,s \\ \hline
			32 & 65 & 3.4\% & 0.33\% & 1.1\,s & 1692\,s \\ \hline
		\end{tabular}
		\label{tab:optgap}
	\end{table}

	\begin{figure}[t]
		\centering
		\definecolor{dc8}{HTML}{6DA7EC}\definecolor{dc16}{HTML}{2A78D6}\definecolor{dc32}{HTML}{184F95}
		\tikzset{lbl/.style={font=\tiny, inner sep=1pt}}
		\begin{tikzpicture}
			\begin{groupplot}[
				group style={group size=1 by 2, vertical sep=1.95cm},
				width=\columnwidth, height=0.56\columnwidth,
				grid=major, grid style={black!10},
				tick label style={font=\scriptsize}, label style={font=\footnotesize}, title style={font=\footnotesize},
				legend style={font=\tiny, fill=white, fill opacity=0.85, text opacity=1, draw=black!30},
				every axis plot/.append style={mark size=1.9pt, line width=0.7pt},
			]
				\nextgroupplot[title={(a) Optimality gap against runtime},
					xmode=log, xmin=0.005, xmax=12000, xtick={0.01,0.1,1,10,100,1000,10000},
					xlabel={Median runtime of a decision (s)},
					ylabel={Additional active PMs (\%)}, ymin=-4, ymax=24,
					legend pos=north east, legend cell align=left]
				\addplot[dc8, mark=*] coordinates {(0.010,21.6) (0.023,10.4) (0.258,3.9)};
				\addlegendentry{8 PMs}
				\addplot[dc16, mark=square*] coordinates {(0.081,14.7) (0.298,5.4) (0.852,4.1)};
				\addlegendentry{16 PMs}
				\addplot[dc32, mark=diamond, mark size=2.4pt] coordinates {(0.289,8.1) (1.104,3.4)};
				\addlegendentry{32 PMs}
				\addplot[black, only marks, mark=o, mark size=2.2pt] coordinates {(0.32,0) (1.7,0) (1692,0)};
				\addlegendentry{OPT}
				\node[lbl, anchor=west, xshift=2pt] at (axis cs:0.010,21.6) {\( M_{\max}{=}2 \)};
				\node[lbl, anchor=east, xshift=-2pt] at (axis cs:0.023,10.4) {4};
				\node[lbl, anchor=north east, xshift=-2pt] at (axis cs:0.258,3.9) {8};
				\node[lbl, anchor=west, xshift=3pt] at (axis cs:0.081,14.7) {4};
				\node[lbl, anchor=west, xshift=3pt] at (axis cs:0.298,5.4) {8};
				\node[lbl, anchor=north, yshift=-2pt] at (axis cs:0.852,4.1) {16};
				\node[lbl, anchor=south, yshift=2pt] at (axis cs:0.289,8.1) {8};
				\node[lbl, anchor=west, xshift=3pt] at (axis cs:1.104,3.4) {16};
				\node[lbl, anchor=north, yshift=-2pt] at (axis cs:0.32,0) {8 PMs};
				\node[lbl, anchor=north, yshift=-2pt] at (axis cs:1.7,0) {16 PMs};
				\node[lbl, anchor=north, yshift=-2pt] at (axis cs:1692,0) {32 PMs};

				\nextgroupplot[title={(b) Runtime as the data center grows},
					xmode=log, log basis x=2, xmin=6.5, xmax=42,
					xtick={8,16,32}, xticklabels={8\\19,16\\30,32\\65}, xticklabel style={align=center},
					xlabel={PMs \( |M| \) (top) and VMs \( |N| \) (bottom)},
					ylabel={Runtime (s)}, ymode=log, ymin=0.04, ymax=20000, ytick={0.1,1,10,100,1000,10000},
					legend pos=north west, legend cell align=left]
				\addplot[black, mark=*, error bars/.cd, y dir=both, y explicit]
					coordinates {(8,0.32) += (0,0.21) -= (0,0.15)
					             (16,1.7) += (0,1.69) -= (0,0.34)
					             (32,1692) += (0,4568) -= (0,1534)};
				\addlegendentry{OPT solve}
				\addplot[dc16, mark=square*, error bars/.cd, y dir=both, y explicit]
					coordinates {(8,0.258) += (0,0.251) -= (0,0.186)
					             (16,0.852) += (0,0.963) -= (0,0.526)
					             (32,1.104) += (0,1.611) -= (0,0.747)};
				\addlegendentry{OrigaMIG call, \( M_{\max}{=}16 \)}
				\addplot[black!45, dotted, line width=0.9pt, domain=8:32, samples=2] {0.32*(x/8)^2};
				\addlegendentry{model size of OPT}
				\addplot[black!45, densely dashed, domain=6.5:42, samples=2, forget plot] {10};
				\node[lbl, text=black!60, anchor=south west] at (axis cs:6.6,10) {10\,s limit of a OrigaMIG call};
				\draw[black!60, <->, >=stealth] (axis cs:30,1.3) -- node[lbl, left, fill=white, fill opacity=0.85, text opacity=1] {\( \approx 1500\times \)} (axis cs:30,1500);
			\end{groupplot}
		\end{tikzpicture}
		\caption{(a) The optimality gap of OrigaMIG against the median runtime of its calls, for each data-center size and neighborhood limit \( M_{\max} \) (labels); OPT has no gap and sits on the \( x \)-axis at its own runtime. A larger neighborhood trades runtime for a smaller gap. (b) Median runtime as the data center grows, with \( M_{\max} = 16 \); the dotted line shows how OPT would grow if its runtime only followed the size of its model. Whiskers in (b) show the 5th to 95th percentile.}
		\label{fig:optgap}
	\end{figure}
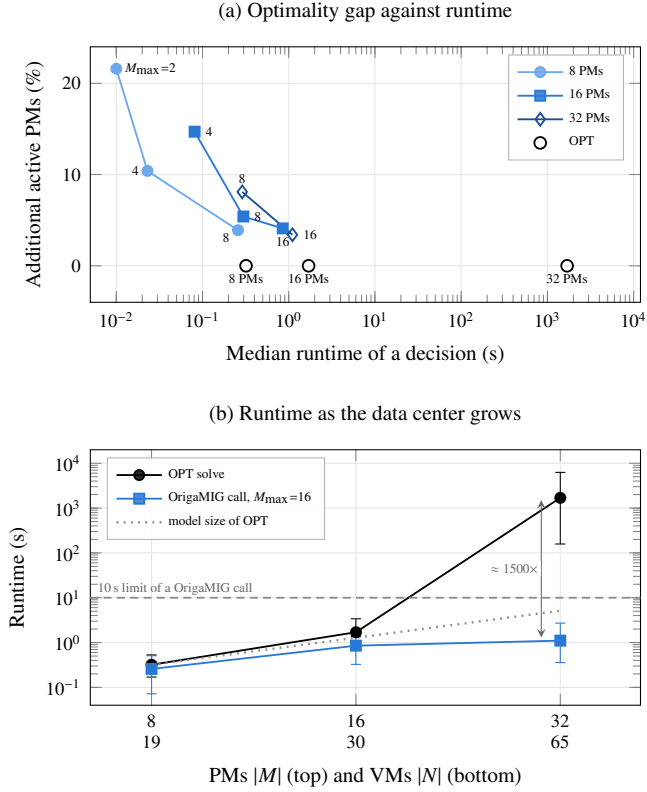

	\emph{Optimality gap.} Table~\ref{tab:optgap} and Fig.~\ref{fig:optgap}(a) show that OrigaMIG uses at most 3--4\% more active PMs than OPT, with no trend in the size of the data center. The relative gap in Eq.~\eqref{eq:1} is at most 0.54\%, because the shared acceptance term dominates the objective. At 8 and 16 PMs a call already covers the whole data center, so the gap stems from when OrigaMIG calls the solver and which PMs it selects, not from the size of the neighborhood. A smaller neighborhood widens the gap. With \( M_{\max} = 8 \), OrigaMIG uses up to 5.4\% more active PMs than OPT at 16 PMs and 8.1\% at 32 PMs, and with \( M_{\max} = 4 \), up to 10.4\% at 8 PMs and 14.7\% at 16 PMs.

	\emph{Runtime.} The size of the restricted BILP is proportional to \( |\hat{N}| (|\hat{M}| + \sum_{j \in \hat{M}} |K_j|) \) (Section~\ref{s:dmig-bilp}), which \( M_{\max} \) and \( N_{\max} \) bound by a constant. The solve time of a call is therefore \( O(1) \) in the size of the data center once it exceeds \( M_{\max} \) PMs, apart from the single pass over the PMs that selects the neighborhood. It varies only with the VMs a neighborhood holds, and its median is 0.9\,s at 16 PMs and 1.1\,s at 32 PMs (Fig.~\ref{fig:optgap}(b)). The model of OPT grows as \( O(|N| (|M| + \sum_{j \in M} |K_j|)) \), and its runtime grows faster than the model, from 0.3\,s at 8 PMs to 28\,min at 32 PMs (Section~\ref{s:bilp-scalability}). These times exceed those of Fig.~\ref{fig:bilp-scaling} at the same number of VMs, because a VM here requests one or two GIs instead of one, which enlarges the BILP. A single solve takes up to 115\,min. At 32 PMs, a OrigaMIG call is about 1500 times faster than OPT, at an optimality gap of 3.4\% more active PMs. For one snapshot at 48 PMs with 94 VMs, OPT took 85\,min and found no better placement than that of OrigaMIG, whose calls took a median of 1.5\,s; for the next snapshots, with 99--105 VMs, OPT did not finish within 16\,h.

	\subsection{Scalability of the BILP}
	\label{s:bilp-scalability}

	We now measure how the time to solve one interval to optimality grows with the number of VMs and the number of PMs. Each PM has two A100 40GB GPUs, 80 cores, and 512 GB of RAM. Each VM requests one GI with a profile drawn uniformly at random from Table~\ref{tab:model-params}, together with 4 cores and 16 GB of RAM per compute slice. The resident VMs are placed by FF. Then, some of them are removed and replaced by new ones to fragment the GPUs, and a batch of new VMs arrives. The weights are those of Section~\ref{s:setup}. CP-SAT runs on eight dedicated CPU cores, starts from the current placement, and stops after \bilpLimitMin{} minutes. Fig.~\ref{fig:bilp-scaling} doubles one of \( |M| \) and \( |N| \) while the other is fixed. Each doubling of \( |N| \) or \( |M| \) approximately doubles the size of the model, but Fig.~\ref{fig:bilp-scaling} shows that the solve time grows much faster. In Fig.~\ref{fig:bilp-scaling}(b), the last two doublings of \( |N| \) multiply the solve time by \bilpNStepA{} and \bilpNStepB{}, and 128 VMs on 64 GPUs already take \bilpNProofOneTwentyEightMin{} minutes. In Fig.~\ref{fig:bilp-scaling}(a), the number of PMs also increases the solve time, even though the added PMs are idle. Beyond about a hundred VMs on a hundred GPUs, the solver proves optimality within the limit in only \bilpBeyondProved{} of \bilpBeyondRuns{} solves. The BILP can therefore be solved to optimality in every interval only for small data centers.

	\section{Discussion}
	\label{s:discussion}

	\emph{Where OrigaMIG helps.} OrigaMIG gains more than TC when the CPU and the RAM of the PMs run out first, with smaller profiles, with either share of tenants, and on the trace (Section~\ref{s:eval-robustness}). Its BILP places the GIs, the cores, and the RAM of a VM together, whereas TC ranks PMs by their GPUs alone, so with fewer cores and RAM, OrigaMIG admits \evalStwoDMIGDaccCpu{} percentage points more than FF while TC admits fewer. TC gains more with larger profiles, with two GIs per VM, and with A30 PMs only. A PM then hosts few VMs, so a neighborhood of 16 PMs holds only 10 to 17 VMs, and the solver proves each call optimal in less than one second. Here the number of PMs per call, and not the time limit, bounds the gains, while TC compacts all GPUs of a class at once. Bounding a neighborhood by its VMs instead of its PMs would let it grow in these cases, and the time limit would still bound each call.

	\emph{Choosing the migration penalty.} The penalty \( \delta_i \) sets how many migrations OrigaMIG spends to save an active PM (Section~\ref{s:eval-migration}). For each period of TC, a value of \( \delta_i \) gives both fewer migrations and fewer active PMs, so an operator can derive \( \delta_i \) from a migration budget instead of tuning a period. The cost of a migration grows with the memory of the VM. On the trace, where most VMs hold a whole GPU, OrigaMIG copies \evalStwoDMIGGbhTrace\,GB per hour, against \evalStwoTCGbhTrace\,GB for TC. Setting \( \delta_i \) in proportion to the memory of VM \( i \) would move large VMs less often.

	\emph{Neighborhood selection.} At a load of 0.85, random neighborhoods keep as few PMs active as Algorithm~\ref{alg:dmig-call} (Section~\ref{s:eval-ablation}). The gains therefore come mainly from solving the BILP on several PMs at once, and a cheaper selection may suffice. Whether this holds at higher loads and in larger data centers remains to be tested.

	\emph{Deployment at scale.} One instance of OrigaMIG starts at most one call per 10\,s step. Its reduction in active PMs over FF falls from \evalSfourScaleDpmOneTwoEight\% with 128 PMs to \evalSfourScaleDpmTwoZeroFourEight\% with 2048 PMs, where it starts \evalSfourScaleCallsTwoZeroFourEight{} of its 360 possible calls per hour. A large data center should therefore run one instance per partition of its PMs, such as a row or a pod, as large-scale allocators do \cite{newell2021ras,narayanan2021pop}, and calls on disjoint neighborhoods can run in parallel (Section~\ref{s:eval-ablation}). A call of a few seconds is short next to a migration between PMs, which takes \evalSthreeDeltaTenCopy\,s on average, and at a load of 1.0, admitted requests wait \evalSoneDMIGWaitOne\,s on average.

	\emph{Limitations.} We evaluate OrigaMIG in simulation. Each run starts from an empty data center, and tenants live 24\,h on average, so the demand reaches 89\% of its steady value when the three measured days start and 96\% on average over them. All policies see the same demand. The energy model switches empty PMs off at no cost and makes the power of a GPU linear in its compute slices in use. A migration copies the memory of the VM once, without further pre-copy rounds or downtime, and GIs do not interfere, as MIG isolates them. The workloads are synthetic, apart from one trace mapped to A100 profiles, and each setting runs three to five seeds. The calls ran on a shared machine, so their runtimes may be longer than on dedicated cores. A deployment on a real cluster, with measured migration times and power, is the next step.

	\section{Conclusion}
	\label{s:conclusion}
	
	This article set out to tackle the challenges introduced by MIG technology in cloud data centers, particularly the fragmentation and inefficiency stemming from rigid placement rules. We showed that the default GI placement policy puts each GI where it maximizes \( \CF \), and that 67\% of the configurations of an A100 arrange their GIs suboptimally. We formulated a BILP model to capture the complexities of acceptance maximization, resource consolidation, and reduced migration overhead. Building on it, we developed OrigaMIG, which places each request with a fast greedy rule and solves the BILP on a small neighborhood of PMs when a request cannot be placed or the GPUs of a class become fragmented. In simulations of 256 GPUs in 128 PMs, OrigaMIG kept fewer PMs active than TC, the stronger of two state-of-the-art MIG-aware policies, in \evalSoneVsTCWins{} of \evalSoneVsTCRuns{} paired runs while migrating \evalSoneVsTCDmigMin\% to \evalSoneVsTCDmigMax\% fewer VMs, and it used the least energy per admitted GPU-hour of all policies. Against the default placement, it kept up to \evalSoneGainDMIGDpmZeropFive\% fewer PMs active and admitted up to \evalSoneGainDMIGDaccOne{} percentage points more of the requested GPU memory. Across eleven variants of the workload and the hardware, it kept fewer PMs active than TC in seven, while TC did better when the VMs were large relative to their GPUs. One instance served up to about 1024 PMs, so larger data centers need one instance per partition of their PMs. On small data centers, OrigaMIG stayed within 4.1\% of the optimal number of active PMs. Future research could extend OrigaMIG by incorporating more advanced prediction models for request patterns, admitting preemptible low-priority VMs into the idle slices that consolidation leaves on active GPUs, or integrating network topology considerations. As MIG becomes increasingly prevalent in modern cloud and HPC deployments, the insights and techniques described in this paper provide a foundation for enhanced GPU resource management in data centers.
	
	\bibliographystyle{elsarticle-num}
	\bibliography{bibliography}
	
\end{document}